\documentclass[prb,twocolumn,showpacs,floatfix]{revtex4-1}

\usepackage[dvipdfmx]{graphicx}
\usepackage{amsmath}
\usepackage{amssymb}
\usepackage{times}
\usepackage{color}
\usepackage{ulem}

\def\a{\alpha}

\def\D{\Delta}

\def\e{\epsilon}

\def\h{\eta}

\def\s{\sigma}

\def\w{\omega}
\def\ua{\uparrow}
\def\da{\downarrow}

\def\Vec#1{\mathbf #1}
\def\ak{a(\Vec{k})}
\def\bk{b(\Vec{k})}
\def\ef{\e_{f_1}}
\def\Vf{V_{1}}
\def\Df{D_{f_1}}
\def\Tc{T_{\rm c}}
\def\DHE{\tilde{D}_{\rm HE}}

\newcommand{\Snor}{\Sigma^{\rm nor}}
\newcommand{\Sano}{\Sigma^{\rm ano}}
\newcommand{\kAN}{\Vec{k}_{\rm AN}}

\begin{document}

\title{Hidden-Fermion Representation of Self-energy in Pseudogap and Superconducting States of Two-Dimensional Hubbard Model}

\author{Shiro Sakai$^1$, Marcello Civelli$^2$, and Masatoshi Imada$^3$}

\affiliation{$^1$Center for Emergent Matter Science, RIKEN, Wako, Saitama 351-0198, Japan\\
$^2$Laboratoire de Physique des Solides, CNRS, Univ. Paris-Sud, Universit\'e Paris-Saclay, 91405 Orsay Cedex, France\\
$^3$Department of Applied Physics, University of Tokyo, Hongo, Tokyo 113-8656, Japan
}
\date{\today}

\begin{abstract}
We study the frequency-dependent structure of electronic self-energy in the pseudogap and superconducting states of the two-dimensional Hubbard model.
We present the self-energy calculated with the cellular dynamical mean-field theory systematically in the space of temperature, electron density, and interaction strength.
We show that the low-frequency part of the self-energy is well represented by a 
simple equation, which describes the transitions of an electron to and from a hidden fermionic state.
By fitting the numerical data with this simple equation, we determine the parameters characterizing 
the hidden fermion and discuss its identity.
The simple expression of the self-energy offers a way to organize numerical data of this uncomprehended
superconducting and pseudogap states, as well as a useful tool to analyze spectroscopic experimental results. The successful description by the simple two-component fermion model supports the idea of ``dark" and ``bright" fermions emerging from a bare electron as bistable excitations in doped Mott insulators.
\end{abstract}
\maketitle

\section{INTRODUCTION}
Electronic correlation effect in solids is considered to be the origin of various interesting electronic properties, however its theoretical description remains an open issue.
In metals, a part of the correlation effect can be described by an increased effective mass of electron, which is then called quasiparticle.
Various properties of weakly-correlated metals are well explained within the quasiparticle picture.
However, as the electronic correlation becomes strong, the weight of the quasiparticle state decreases and is transferred to high-energy states.
Virtual transitions of an electron to and from other states produce a frequency-dependent electron self-energy.
If strong correlation effects produce an emergent electronic state at low energy, the virtual transitions to and from this state may give a singular self-energy at low energy, which cannot be described only by the effective mass. This singularity of the self-energy can then yield various 
anomalies in the electronic properties of the metal.

High-temperature superconducting cuprates are a prototype of the strongly-correlated electron systems.
In fact, the superconductivity emerges when carriers are doped into the strongly-correlated Mott-insulating state of the mother compound.  
Various anomalous behaviors have been observed in spectroscopic experiments both for the superconducting phase and for the normal phase above the superconducting transition temperature $\Tc$. 
The understanding of the low-energy electron dynamics is crucial to understand these anomalous behaviors and ultimately the mechanism of the high-temperature superconductivity.

The dynamical mean-field theory \cite{metzner89,georges96} and its extensions are suitable tools to numerically calculate the frequency-dependent properties in strongly correlated systems. 
In fact, the cellular dynamical mean-field theory (CDMFT) \cite{kotliar01}, dynamical cluster approximation \cite{maier05RMP}, and related quantum cluster theories \cite{senechal00,potthoff03} have been successfully applied to the two-dimensional (2D) Hubbard model, the standard model for the cuprates.
Much effort has been put into the clarification of the electron dynamics in the $d$-wave superconducting phase\cite{lichtenstein00,maier05PRL,capone06,aichhorn07,haule07,kancharla08,civelli08,kyung09,civelli09PRL,civelli09PRB,sordi12PRL,gull13PRL,gull15,sakai16,harland16} and in the anomalous-metal pseudogap phase \cite{senechal00,huscroft01,maier02,senechal04,civelli05,kyung06PRB1,stanescu06,
liebsch09,sakai09PRL,sakai10,okamoto10PRB82,eder11,kohno12,sordi12PRL,sakai13,kohno14,merino16,yang16}.
These studies have revealed the presence of a sharp low-energy peak in the self-energy of both the normal \cite{maier02,stanescu06,kyung06PRB1,sakai09PRL,liebsch09,sakai10,eder11} and superconducting states \cite{haule07,maier08,kyung09,civelli09PRL,gull15,sakai16}.
The self-energy peak in the normal state yields the pseudogap in the spectral function, whereas the peak in the superconducting state strengthens the superconductivity considerably and is at the origin of the high $\Tc$ \cite{maier08}.

In the previous work \cite{sakai16}, we showed that these self-energy peaks above and below $\Tc$ are smoothly connected with each other as the temperature changes across $\Tc$.
Below $\Tc$, we found that the self-energy peaks of the normal component cancel with those of the anomalous component in the single-particle Green's function. This cancellation of the self-energy peaks leads to three significant consequences: 
First, the prominent peak in the anomalous self-energy, triggering the high $\Tc$, becomes hidden in conventional experimental tools that measure single-particle spectra.
Second, the self-energy peaks are smoothly connected to {\it poles} at zero temperature, 
which represent a fermionic excitation hybridizing with the low-energy electron.
Third, the superconducting gap and the pseudogap in the spectral function involve different singularities since only the latter is directly generated by the self-energy peak.
Based on the second observation, we constructed a simple phenomenological model consisting of the low-energy electron and a hidden fermion which is to be integrated out to discuss the electron dynamics.
We showed that the model indeed describes well the low-energy electron dynamics in the pseudogap and superconducting states of the 2D Hubbard model \cite{sakai16}.

The phenomenological model is a solvable one-body model and provides us with a useful insight
into the strong-correlation effect.
The derived simple expression of the self-energy will also be useful in analyzing experimental results.
In this article, we first introduce multiple hidden fermions on the basis of Luttinger's spectral respresentation of the self-energy \cite{luttinger61}, and derive a general hidden-fermion representation of the electronic self-energy.
We then focus on the pseudogap and superconducting states in the 2D Hubbard model, where only one hidden fermion is relevant at low energies.
Using the CDMFT solved with the exact-diagonalization (ED) method, we calculate a precise real-frequency structure of the self-energy and present a thorough investigation of the parameter dependence of the sructure.  
We analyze the CDMFT self-energy based on this hidden-fermion representation, and study how the phenomenological-model parameters change with doping concentration, temperature, and the Hubbard interaction $U$.
These dependences in turn characterize the low-energy hidden fermion, providing a clue to identify it.

This paper is organized as follows.
In Sec.~\ref{sec:method}, we introduce the Hubbard Hamiltonian, the CDMFT, and the hidden-fermion model. We derive a general expression of the self-energy in the hidden-fermion model (the detailed derivation is given in Appendix A), and then derive a simple but approximate expression for the pseudogap and supercondcting states of the 2D Hubbard, based on the fact that the self-energy in these states involves essentially only one pole at low energy. 
In Sec.~\ref{sec:result}, we present the numerical results obtained with the CDMFT.
The results are interpreted in terms of the hidden fermion in Sec.~\ref{sec:interpret}, where we determine the parameters in the hidden-fermion model by fitting the CDMFT self-energy.
Sec.~\ref{sec:discuss} is devoted to discussions on the results obtained in the previous sections.
There we also discuss the relation between the multiple-hidden-fermion model and the two-component fermion model introduced in Ref.~\onlinecite{sakai16}, and clarify the meaning of the ``bright'' and ``dark'' fermions in the abstract.
We summarize the paper in Sec.~\ref{sec:summary}.

\section{MODEL AND METHOD} \label{sec:method}
\subsection{2D Hubbard model} \label{ssec:HM}
We study the 2D single-band Hubbard model on a square lattice. The Hamiltonian reads
\begin{align}
H_\text{Hubbard}= \sum_{\Vec{k}\s}\e_c(\Vec{k})c_{\Vec{k}\s}^\dagger c_{\Vec{k}\s}^{\phantom {\dagger}}
+ U\sum_{i}n_{i\ua}n_{i\da},
\label{hubbard}
\end{align}
where $c_{\Vec{k}\s}^{\phantom {\dagger}}$ $(c_{\Vec{k}\s}^\dagger)$ annihilates (creates) an electron with spin $\s$ and momentum $\Vec{k} = (k_x, k_y)$, 
$c_{i\s}^{\phantom {\dagger}}$ $(c_{i\s}^\dagger)$ is its Fourier component at site $i$, and
$n_{i\s}\equiv c_{i\s}^\dagger c_{i\s}^{\phantom {\dagger}}$.
\begin{align}
\e_c(\Vec{k})\equiv -2t(\cos k_x + \cos k_y) -4t' \cos k_x\cos k_y -\mu
\label{disp}
\end{align}
is the bare dispersion with $t$ $(t')$ being the (next-)nearest-neighbor transfer integral and $\mu$ being the chemical potential. $U$ represents the onsite Coulomb repulsion.
We use $t=1$ as the unit of energy. For hole-doped cuprates, the energy scale is roughly given by $t=1\sim0.3$eV\cite{hybertsen90}. 
We fix $t'=-0.2t$ and concentrate on the dependence on the temperature $T$, the electron density $n$, and the interaction strength $U$. 
We define the doping $\delta= 1-n$ with respect to the parent Mott insulator at $n=1$. 
For the self-energy peaks of our interest, we expect qualitatively the same results for $-0.4\lesssim t' \lesssim 0$, 
as we have indeed confirmed for several sets of the other parameters. 
The self-energy peaks are in fact a consequence of a proximity to the Mott insulator rather than the itinerant character of the electrons.

\subsection{CDMFT}
Within the CDMFT \cite{kotliar01}, we map the model (\ref{hubbard}) onto an effective impurity problem consisting of an interacting 2$\times$2 
cluster coupling to a dynamical (frequency-dependent) mean field $\hat{g}_0(\omega)$. 
We solve the effective impurity problem with a finite-temperature extension of the ED method \cite{capone07,liebsch08,liebsch12}.
An advantage of the ED is that we can directly calculate Green's functions and self-energies on the {\it real-frequency} axis, without any additional analytic continuation scheme used for example in quantum Monte-Carlo solvers. This is crucially important for our purpose of studying the precise frequency-dependent electronic structures.
We instead sacrifice a resolution in momentum space: The clusters beyond 2$\times$2 are intractable with the ED.
We therefore focus on the frequency dependence rather than the momentum dependence.  

When we solve the impurity problem with ED, $\hat{g}_0$ is replaced with an approximate function, $\hat{g}_0^\text{ED}$,  generated from eight noninteracting bath sites. 
The bath-site parameters are determined by minimizing the distance function 
\begin{align}
d=\sum_{i,j,n} \frac{1}{\w_n} \left | \left[\hat{g}_0(i\w_n)^{-1} - \hat{g}_0^\text{ED}(i\w_n)^{-1} \right]_{ij}\right | .
\end{align}
The factor $1/\w_n$ is inserted in order to obtain reliable results at low-energy, which are of our primary interest,
while we somewhat sacrifice the accuracy for high-energy features.
For the latter, we avoid a detailed analysis and discuss only the most promient structure, which is related to the Mott gap. We have checked that our conclusion does not change within moderate choices of $d$.
  
Since, in the superconducting state, the number of electrons is not conserved, we block-diagonalize the effective impurity Hamiltonian with respect to the $z$ component $S_z$ of the total spin; the maximal block ($S_z=0$) for 12 sites has 2,704,156 dimensions, which is much larger than the maximal block in the normal state (853,776 dimensions for the $n_\ua=n_\da=6$ block when we categorize the Hilbert space with respect to the number of electrons with each spin, $n_\ua$ and $n_\da$) for the same system size.
The self-consistent loops of the CDMFT are performed with the quantities defined at the Matsubara frequencies $\w_n=(2n-1)\pi T$ ($n=0,\pm1,\pm2,\cdots$).

After a convergence is reached, we calculate Green's function at real frequencies by replacing $i\w_n$ with $\w+i\h$, 
where $\h(\w)=\h_0 \min\{5,1+\w^2\}$ with $\h_0=0.05$ is the smearing factor.
We adopt the frequency-dependent form \cite{liebsch12,civelli09PRB} because the small value of $\h$ 
at low energy allows us to resolve the precise self-energy structure while the relatively large value of $\h$ smoothens the shaggy high-energy structure which could depend on the initial arrangement of the bath sites.
Although the smearing factor changes the sharpness of peaks, the positions and weights of the peaks are insensitive to moderate choices of $\h$.
In the following equations for real-frequency properties, $\w$ should be read as $\w+i\h$ (except for the one in the arguments of a function).

A typical output of the CDMFT is the single-particle Green's function. In the superconducting state, it is written in the Nambu-matrix form as
\begin{align}
 \hat{G}(\Vec{k},\w)=\left(
    \begin{array}{cc}
     G(\Vec{k},\w) & F(\Vec{k},\w)\\
     F(\Vec{k},\w) & G(\Vec{k},-\w)^\ast
    \end{array}
   \right),
\label{G}
\end{align}
where $G$ ($F$) is the retarded normal (anomalous) Green's function. 
Here we consider a spin-singlet superconducting order, for which we can choose
the $U(1)$ gauge to make the two offdiagonal anomalous components to be the same. 
The spectral function is defined by
\begin{align}
 A(\Vec{k},\w)\equiv -\frac{1}{\pi}\text{Im}G(\Vec{k},\w).
\label{eq:akw}
\end{align}
The normal and anomalous self-energies, $\Snor$ and $\Sano$, are defined by 
\begin{align}
 &\left(
    \begin{array}{cc}
     \Snor(\Vec{k},\w) & \Sano(\Vec{k},\w)\\
     \Sano(\Vec{k},\w) & \Snor(\Vec{k},-\w)^\ast
    \end{array}
   \right) \nonumber \\
   &=
    \left(
    \begin{array}{cc}
     \w-\e_c(\Vec{k}) & 0\\
     0 & \w+\e_c(\Vec{k})
    \end{array}
   \right) - \left[\hat{G}(\Vec{k},\w)\right]^{-1}.
\label{sig}
\end{align}

With the 2$\times$2 cluster, we obtain these quantities at essentially four momentum points; (0,0), ($\pi$,0), (0,$\pi$) and ($\pi$,$\pi$). 
By symmetry, the quantities at ($\pi$,0) and (0,$\pi$) are equivalent except for the overall sign of the anomalous part.
The 2$\times$2 CDMFT in fact outputs only the nearest-neighbor component for $\Sano$ (while the local and next-nearest-neighbor components are numerically zero), 
displaying $x$ and $y$ real-space components, and correspondingly ($\pi$,0) and (0,$\pi$) momentum components, with opposite signs. As a consequence, the
superconducting gap acquires a $d$-wave symmetry and the (0,0) and ($\pi$,$\pi$) components are vanishing on the so-called nodal line.
For these reasons, we concentrate on the quantities at $\Vec{k}=\kAN\equiv (\pi,0)$ in Sec.~\ref{sec:result}.

\subsection{Hidden Fermion model}\label{ssec:tcfm}
As we shall show below, the CDMFT results are well represented by a phenomenological one-body model where an electron transfers to and from various hidden fermionic states, which are integrated out to give a self-energy to the electron.
The phenomenological model is useful to understand our numerical results and offers us a deep insight into the mechanism of the superconductivity and the pseudogap, as well as their relationship.
The simple equation, Eq.~(\ref{sigc_sc}) or (\ref{sigc_sc1he}) below, for the self-energy will also be useful to analyze experimental results by taking into account the strong correlation effects.

\begin{figure}[tb]
\center{\includegraphics[width=0.4\textwidth]{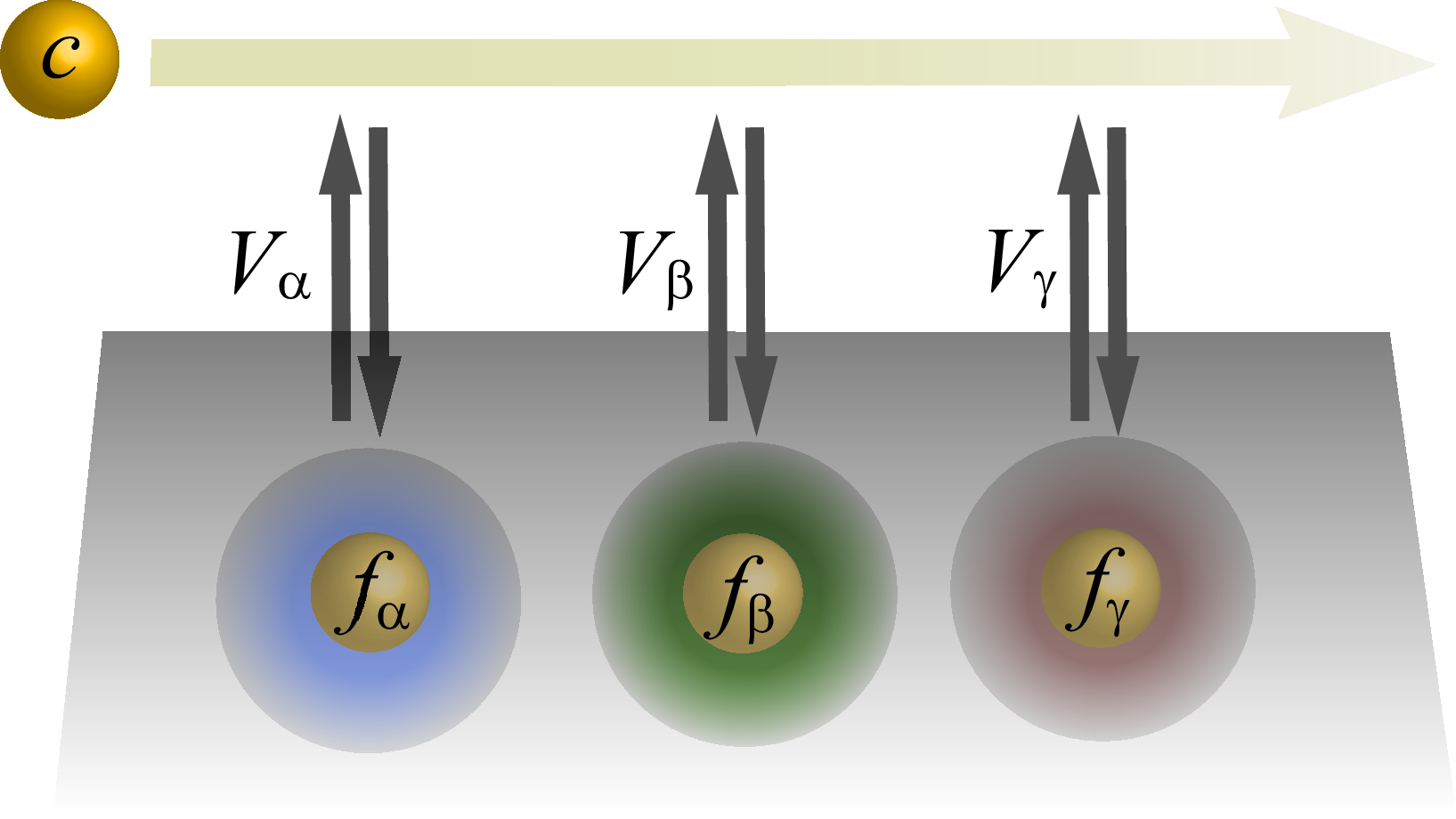}}
\caption{(Color online). Schematic illustration of the model (\ref{tcf_ns}). An electron propagates the crystal, with sometimes transferring to virtual states $f_\a$ through the hybridization $V_\a$.}
\label{fig:hidden}
\end{figure}

\subsubsection{Normal state}\label{sssec:ns}
We start with the normal state \footnote{This part for the normal state is essentially equivalent to what has been done in a recent paper Ref.~\onlinecite{seki16}.}.
In general, the imaginary part of $\Snor$ can be represented by a superposition of many $\delta$-functional peaks. Then, considering the analytic property, we can write it in the form
\begin{align}
 \Snor(\Vec{k},\w)=s(\Vec{k})+\sum_{\a=1,2,\dots} \frac{V_\a(\Vec{k})^2}{\w-\e_{f_\a}(\Vec{k})},
\label{signor_ns}
\end{align}
where $s(\Vec{k})$ is the $\w$-independent part of the self-energy \cite{luttinger61}.
The second term can describe a singular structure as an isolated pole, and a nonsingular structure 
as a continuum of poles.
Note that in the numerical simulation of a finite-size system, we use a broadening factor $i\eta$ for $\w$. 

Equation (\ref{signor_ns}) is equivalent to the self-energy of electron $c$ in the following Hamiltonian,
\begin{align}
H_{cf,\text{normal}}= \sum_{\Vec{k}\s} & 
          \left[ \{\e_c(\Vec{k})+s(\Vec{k})\} c_{\Vec{k}\s}^\dagger c_{\Vec{k}\s}^{\phantom {\dagger}} \right.\nonumber\\
         &+ \sum_{\a}\e_{f_\a}(\Vec{k}) f_{\a\Vec{k}\s}^\dagger f_{\a\Vec{k}\s}^{\phantom {\dagger}} \nonumber \\
         & \left. +\sum_{\a}V_\a(\Vec{k})( f_{\a\Vec{k}\s}^\dagger c_{\Vec{k}\s}^{\phantom {\dagger}} 
                                       +c_{\Vec{k}\s}^\dagger f_{\a\Vec{k}\s}^{\phantom {\dagger}} ) 
              \right],
\label{tcf_ns}
\end{align}
when $f$ degrees of freedom are integrated out in the path-integral formalism of the corresponding action $S_{cf,\text{normal}}$.
Namely, by defining an effective action $S_\text{normal}^\text{eff}$ of only $c$ degrees of freedom through 
\begin{align}
&\exp\left(-S_{\text{normal}}^\text{eff}[c^\dagger,c]\right)\nonumber\\
\propto \ &
\int \mathcal{D}f^\dagger \mathcal{D}f \exp\left(-S_{cf,\text{normal}}[c^\dagger,c,f^\dagger,f]\right) ,
\label{Seff_ns}
\end{align}
we obtain the self-energy correction given by Eq.~(\ref{signor_ns}) for the $c$ fermion  (see Appendix A for derivation; another proof can be found in a recent paper, Ref.~\onlinecite{seki16}). Here we have abbreviated the indices in $c$ and $f$.
In Eq.~(\ref{tcf_ns}),  $\e_c(\Vec{k})$ and $\e_{f_\a}(\Vec{k})$ represent the bare dispersions of  the $c$ and $f_{\a}$ fermions, respectively, and $V_\a(\Vec{k})$ represents a hybridization between them.

In the following, we sometimes omit the subscript or argument of $\Vec{k}$ just for brevity.
Since the model (\ref{tcf_ns}) is diagonal in $\Vec{k}$, the $\Vec{k}$ dependence can always be recovered immediately.
We illustrate the model (\ref{tcf_ns}) in Fig.~\ref{fig:hidden}.

\subsubsection{Superconducting state}
We extend the Hamiltonian (\ref{tcf_ns}) to the superconducting state, by introducing anomalous terms as
\begin{align}
H_{cf,\text{super}}=& H_{cf,\text{normal}} 
                      -\sum_{\Vec{k}} D_c(\Vec{k}) (c_{\Vec{k}\ua} c_{-\Vec{k}\da} +\text{H. c.})\nonumber\\ 
                   -& \sum_{\a\Vec{k}} D_{f_\a}(\Vec{k}) (f_{\a\Vec{k}\ua} f_{\a-\Vec{k}\da} +\text{H. c.}) .
\label{tcf_sc}
\end{align}
Similarly to the case of the normal state, integrating out the $f$ degrees of freedom in Eq.~(\ref{tcf_sc}) yields the self-energy,
\begin{align}
 \Snor(\Vec{k},\w)&=s(\Vec{k})+\sum_\a\frac{V_\a(\Vec{k})^2[\w+\e_{f_\a}(\Vec{k})]}{\w^2-\e_{f_\a}(\Vec{k})^2-D_{f_\a}(\Vec{k})^2},\nonumber\\
 \Sano(\Vec{k},\w)&=D_c(\Vec{k})+\sum_\a\frac{-V_\a(\Vec{k})^2 D_{f_\a}(\Vec{k})}{\w^2-\e_{f_\a}(\Vec{k})^2-D_{f_\a}(\Vec{k})^2}
\label{sigc_sc}
\end{align}
(see Appendix A for derivation). These equations show that both $\Snor$ and $\Sano$ have poles at the same electron-hole symmetric energies, $\w=\pm\sqrt{\e_{f_\a}^2+D_{f_\a}^2}$ due to the presence 
of $V_\a$ and $D_{f_\a}$, while $D_c$ does not produce any frequency-dependent structure.

\subsubsection{General remarks for Hubbard model}\label{sssec:case_hubbard}
We apply the general equations, (\ref{signor_ns}) and (\ref{sigc_sc}),  to the normal state of the Hubbard model. 
In this case, as shown in Ref.~\onlinecite{eder11}, the analysis of the first-order moment of the normal Green's function gives 
\begin{align}
s(\Vec{k}) = \frac{n}{2}U. 
\label{sk}
\end{align}
Moreover, in the normal state, the analysis of the second-order moment gives a sum rule,
\begin{align}
\sum_\a V_\a(\Vec{k})^2 =  \frac{n}{2} \left(1-\frac{n}{2}\right)U,
\label{sumrule_ns}
\end{align}
at each $\Vec{k}$, as proved in Ref.~\onlinecite{seki11}.
This sum rule can be extended to the superconducting state as
\begin{align}
\sum_\a V_\a(\Vec{k})^2 +D_c(\Vec{k})^2= \frac{n}{2} \left(1-\frac{n}{2}\right)U,
\label{sumrule_sc}
\end{align}
as derived in Appendix B.

\subsubsection{Underdoped regime in Hubbard model}\label{sssec:single}

For the pseudogap state in the 2D Hubbard model, many previous studies \cite{kyung06PRB1,stanescu06,liebsch09,sakai09PRL,sakai09PhysB,sakai10,eder11,sakai13,sakai16} have shown that essentially one pole in Eq.~(\ref{signor_ns}) is relevant at low energy.
In terms of Eq.~(\ref{tcf_ns}), this means that only one $f$ fermion, say $f_1$, is present at low energy, being isolated from other $f$ fermions at higher energies. Namely,
\begin{align}
 \Snor(\Vec{k},\w)\simeq \frac{n}{2}U +\frac{V_1(\Vec{k})^2}{\w-\e_{f_1}(\Vec{k})}.
\label{signor_ns1}
\end{align} 
The isolated pole in $\Snor$ at $\w=\e_{f_1}$ generates the pseudogap in the spectral function.

This isolated pole persists even in the superconducting state \cite{gull15,sakai16} of the 2D Hubbard model, giving an electron-hole symmetric pair of poles in $\Sano$. The latter poles have been reported previously \cite{haule07,maier08,kyung09,civelli09PRL,gull13PRL,gull15,sakai16}. 
According to Eq.~(\ref{sigc_sc}), this leads to 
\begin{align}
 \Snor(\Vec{k},\w)&\simeq \frac{n}{2}U+\frac{V_1(\Vec{k})^2[\w+\e_{f_1}(\Vec{k})]}{\w^2-\e_{f_1}(\Vec{k})^2-D_{f_1}(\Vec{k})^2},\nonumber\\
 \Sano(\Vec{k},\w)&\simeq D_c(\Vec{k})+\frac{-V_1(\Vec{k})^2 D_{f_1}(\Vec{k})}{\w^2-\e_{f_1}(\Vec{k})^2-D_{f_1}(\Vec{k})^2}
\label{sigc_sc1}
\end{align}
Notice that finite values of $D_{f_1}$ and $V_1$, rather than the static $D_c$ term, yield the poles in $\Sano$.  

An interesting property of Eq.~(\ref{sigc_sc1}) is that, in Green's functions $G$ and $F$, the poles of $\Snor$ cancel with those of the anomalous contribution, as far as the poles of $f_1$ are energetically well isolated from others.
The cancellation of the poles is indeed seen in the CDMFT results.
While it would not be so easy in general to numerically distinguish a single-pole-like peak from other peaks (because of the usage of the broadening factor $\eta$), the cancellation qualitatively evidenced the isolated nature of the pole in the CDMFT self-energy \cite{sakai16}.

\subsubsection{High-energy corrections}\label{sssec:he}
In order to pursue a quantitative reproduction of the CDMFT self-energies with simple expressions, we further improve Eqs.~(\ref{signor_ns1}) and (\ref{sigc_sc1}) by adding the lowest-order corrections due to contributions from high-energy poles. 
In Eq.~(\ref{signor_ns}), when $|\e_{f_\a}|$ ($\a=2,3,\dots$) is large compared to $|\w|$, we can expand the high-energy contribution in $\frac{\w}{\e_{f_\a}}$ as
\begin{align}
 &    \sum_{\a=2,3,\dots} \frac{V_\a^2}{\w-\e_{f_\a}}\nonumber\\
 &=-\sum_{\a=2,3,\dots} \frac{V_\a^2}{\e_{f_\a}}\left[ 1+\frac{\w}{\e_{f_\a}}+O\left( \left(\frac{\w}{\e_{f_\a}}\right)^2\right) \right] \nonumber\\
 &\simeq \ak-\bk\w.
\label{signor_he}
\end{align}
Here, $\ak\equiv-\sum_{\a=2,3,\dots} \frac{V_\a(\Vec{k})^2}{\e_{f_\a}(\Vec{k})}$ gives an $\w$-independent shift, which can be absorbed into the one-body term of the $c$ fermion in model (\ref{tcf_ns}), and $\bk\equiv\sum_{\a=2,3,\dots} \left[ \frac{V_\a(\Vec{k})}{\e_{f_\a}(\Vec{k})}\right] ^2$ contributes to the band renormalization of $c$.
Combining Eqs.~(\ref{signor_ns}), (\ref{sk}) and (\ref{signor_he}), we obtain 
\begin{align}
 \Snor(\Vec{k},\w)\simeq\frac{V_1(\Vec{k})^2}{\w-\e_{f_1}(\Vec{k})} +\tilde{a}(\Vec{k})-\bk\w
\label{signor_ns1he}
\end{align}
with $\tilde{a}(\Vec{k})\equiv\ak+\frac{n}{2}U$, for the low-energy part.

In the superconducting state, the poles are located at $\w=\pm\sqrt{\e_{f_\a}^2+D_{f_\a}^2}$ [see Eq.~(\ref{sigc_sc})].
Then, for $|\w|\ll \sqrt{\e_{f_\a}^2+D_{f_\a}^2} (\a=2,3,\cdots)$, the contribution from the high-energy poles is written as
\begin{align}
 \sum_{\a=2,3,\cdots}\frac{V_\a^2 (\w+\e_{f_\a})}{\w^2-\e_{f_\a}^2-D_{f_\a}^2}&= 
 \ak -\bk \w + O\left(\frac{\w^2}{\e_f^2+D_f^2}\right),\nonumber\\
 \sum_{\a=2,3,\cdots} \frac{V_\a^2 D_{f_\a}}{\w^2-\e_{f_\a}^2-D_{f_\a}^2}
 &= D_{\rm HE}(\Vec{k})+ O\left(\frac{\w^2}{\e_f^2+D_f^2}\right)
\label{sigc_sc_he}
\end{align}
with $\ak\equiv -\sum_{\a=2,3,\cdots}\frac{V_\a^2 \e_{f_\a}}{\e_{f_\a}^2+D_{f_\a}^2}$, $\bk\equiv \sum_{\a=2,3,\cdots}\frac{V_\a^2}{\e_{f_\a}^2+D_{f_\a}^2}$, and 
$D_{\rm HE}(\Vec{k})\equiv \sum_{\a=2,3,\cdots} \frac{V_\a^2 D_{f_\a}}{\e_{f_\a}^2+D_{f_\a}^2}$: $\ak$ and $\bk$ here for the superconducting state are natural extentions of those for the normal state so that we use the same notations.
Combining Eqs.~(\ref{sigc_sc}), (\ref{sk}) and (\ref{sigc_sc_he}), we obtain 
\begin{align}
 \Snor(\Vec{k},\w)&\simeq \frac{V_1(\Vec{k})^2(\w+\e_{f_1}(\Vec{k}))}{\w^2-\e_{f_1}(\Vec{k})^2-D_{f_1}(\Vec{k})^2} 
 + \tilde{a}(\Vec{k})-\bk\w,\nonumber\\
 \Sano(\Vec{k},\w)&\simeq \frac{-V_1(\Vec{k})^2 D_{f_1}(\Vec{k})}{\w^2-\e_{f_1}(\Vec{k})^2-D_{f_1}(\Vec{k})^2}+\tilde{D}_{\rm HE}(\Vec{k})
\label{sigc_sc1he}
\end{align}
at low energy. $\tilde{D}_{\rm HE}(\Vec{k})=D_{\rm HE}(\Vec{k})+D_c(\Vec{k})$ can be absorbed into the anomalous term of $c$ in the model (\ref{tcf_sc}). Notice that for $D_c=D_{f_\a}=0$, the above equations reduce to Eq.~(\ref{signor_ns1he}).

\subsubsection{Fitting procedure}\label{sssec:fit}
In Sec.~\ref{sec:interpret}, we shall determine the model parameters ($\e_{f_1}$, $D_{f_1}$, $V_1$, $\tilde{a}$, $b$ and $\tilde{D}_{\rm HE}$) at $\Vec{k}=(\pi,0)$ in Eqs.~(\ref{signor_ns1he}) and (\ref{sigc_sc1he}) by a least-square fitting of the CDMFT self-energies.
As one can easily see in an identity,
\begin{align}
 \frac{1}{\w+i\eta-\e}=\frac{\w-\e}{(\w-\e)^2+\eta^2}-i\frac{\eta}{(\w-\e)^2+\eta^2}
\label{1/w}
\end{align}
(where we have explicitly written the broadening factor $\eta$), a pole (at $\w=\e-i\h$ here) influences 
a wider energy range in the real part than in the imaginary part.
Therefore, in order to disentangle the contribution from each pole in the numerical result, 
the imaginary part is more tractable than the real part. 

For this reason, we first look at the imaginary part of the CDMFT self-energy and determine
the fitting parameters, $\e_{f_1}$, $D_{f_1}$ and $V_1$, which are related to the low-energy pole.
Notice that $\tilde{a}$, $b$ and $\tilde{D}_{\rm HE}$ are irrelevant to this fitting of the imaginary part
 since they are real.
Then, we look at the real part of the CDMFT self-energy, to determine $\tilde{a}$, $b$ and $\tilde{D}_{\rm HE}$.
This is done again with the least-square fitting. The parameter obtained with this procedure will
be plotted in Sec.~\ref{sec:interpret}.

\section{CDMFT RESULTS}\label{sec:result}

In this section, we present the results obtained with the CDMFT calculations.
We clarify in Sec.~\ref{ssec:dop} the parameter regions which we study. 
We comprehensively study the evolusion of the electronic structure with temperature (Sec.~\ref{ssec:akw_tdep}), hole doping (Sec.~\ref{ssec:akw_ndep}), and Hubbard $U$ (Sec.~\ref{ssec:akw_udep}).
While our main interest is in the low-energy electronic structure, we present the electronic structure in a global energy range, too, because they may be correlated with each other.

We focus on the frequency-dependent structure mainly at $\Vec{k}=\kAN=(\pi,0)$.
This is because the present 2$\times$2 CDMFT does not have a high momentum resolusion and is most reliable at the cluster momenta $(0,0)$, $(\pm\pi,\pm\pi)$, $(\pm\pi,0)$, $(0,\pm\pi)$, among which $(\pi,0)$ [or equivalently $(-\pi,0)$ and $(0,\pm\pi)$] is of the most interest: it is close to the normal-state Fermi surface and the $d$-wave superconducting gap is maximized around the point.
Note that the superconducting gap vanishes at $(0,0)$ and $(\pm\pi,\pm\pi)$. 
While a similar behavior is reasonably expected in a region close to $(\pi,0)$, it would also be interesting to study the momentum region around the nodal point, $(\frac{\pi}{2},\frac{\pi}{2})$. This however requires a larger cluster calculation intractable at present with the ED solver and remains a future issue.  

\subsection{Superconducting order parameter}\label{ssec:dop}
\begin{figure}[tb]
\center{
\includegraphics[width=0.48\textwidth]{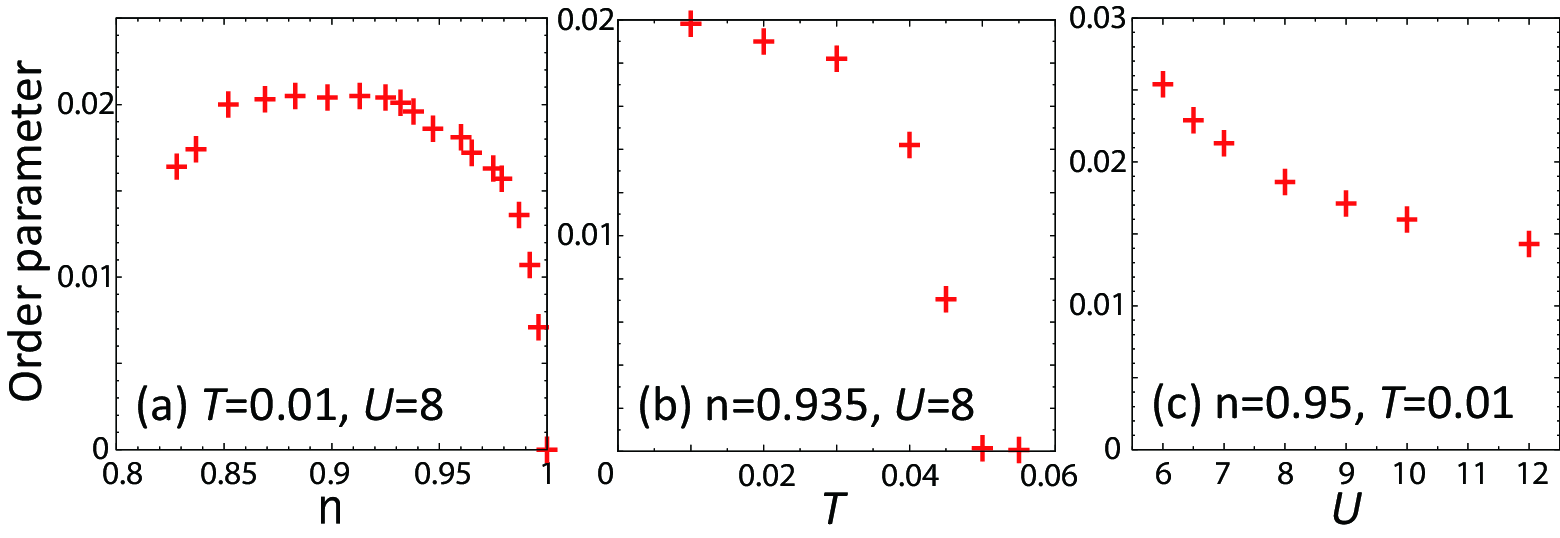}}
\caption{(Color online). $d$-wave superconducting order parameter plotted against (a) $n$ at $T=0.01$ and $U=8$, (b) $T$ for $n=0.935$ and $U=8$, and (c) $U$ for $n=0.95$ and $T=0.01$.
The unit of energy is given by $t=1$.}
\label{fig:dop}
\end{figure}

In order to clarify the parameter region studied in the subsequent sections, in Fig.~\ref{fig:dop} we show several plots of  the $d$-wave superconducting order parameter, $\langle c_{i\uparrow}c_{j\downarrow}\rangle$, with $i$ and $j$ being the nearest-neighbor sites.
Panel (a) plots it against the electron density $n$ at $T=0.01$ and $U=8$.
We see that the order parameter is maximal at around $n=0.88-0.92$ (which we call optimal doping) and decreases for both lower $n$ (overdoped region) and higher $n$ (underdoped region).
At $n=1$ the system becomes a Mott insulator and the superconducting order parameter vanishes.
The result is consistent with the previous studies at zero temperature \cite{kancharla08,civelli09PRL}.
The optimal doping is somewhat shifted to a high density compared to that ($\sim0.85$) of the real cuprates, likely because of the smallness of the used cluster size (i.e., 2$\times$2).
This quantitative difference of the optimal doping however will not affect our conclusions as far as we follow the above definitions of the optimal, overdoped, and underdoped regions.

Panel (b) plots the order parameter against $T$ for $n=0.935$ and $U=8$.
We find that $\Tc$ is between $T=0.045$ and 0.05. 
As a rough estimate, for a typical value of $t\sim 0.3$~eV $\sim$ 3,000~K, this $\Tc$ is evaluated to be 135 - 150 K, which is a bit too high but still in a reasonable agreement with the experimental $\Tc$ of single-layer cuprates.

Panel (c) plots the order parameter against $U$ for $n=0.95$ and $T=0.01$.
We have restricted the plot range to $U \geq 6$, for which the Mott-insulating state appears at $n=1$ \cite{zhang07,park08} and hence the Mott physics is relevant for $n<1$. 
As a matter of fact, our 2$\times$2 CDMFT is expected to be more valid for the strongly-correlated regime where the electrons are more localized.
We see that the order parameter decreases monotonically with $U$. 
The decay is slower than $t^2/U$ for $U>8$.
Aside from Sec.~\ref{ssec:akw_udep}, we use $U=8$, which is considered to be reasonable for hole-doped cuprates.
Note also that in our studies on the $T$ dependence and on the $U$ dependence, we focus on the underdoped region ($n=0.935$ and $0.95$, respectively), where the nontrivial self-energy pole of our interest is most pronounced.

\subsection{Temperature dependence}\label{ssec:akw_tdep}

\begin{figure}[tb]
\center{\includegraphics[width=0.48\textwidth]{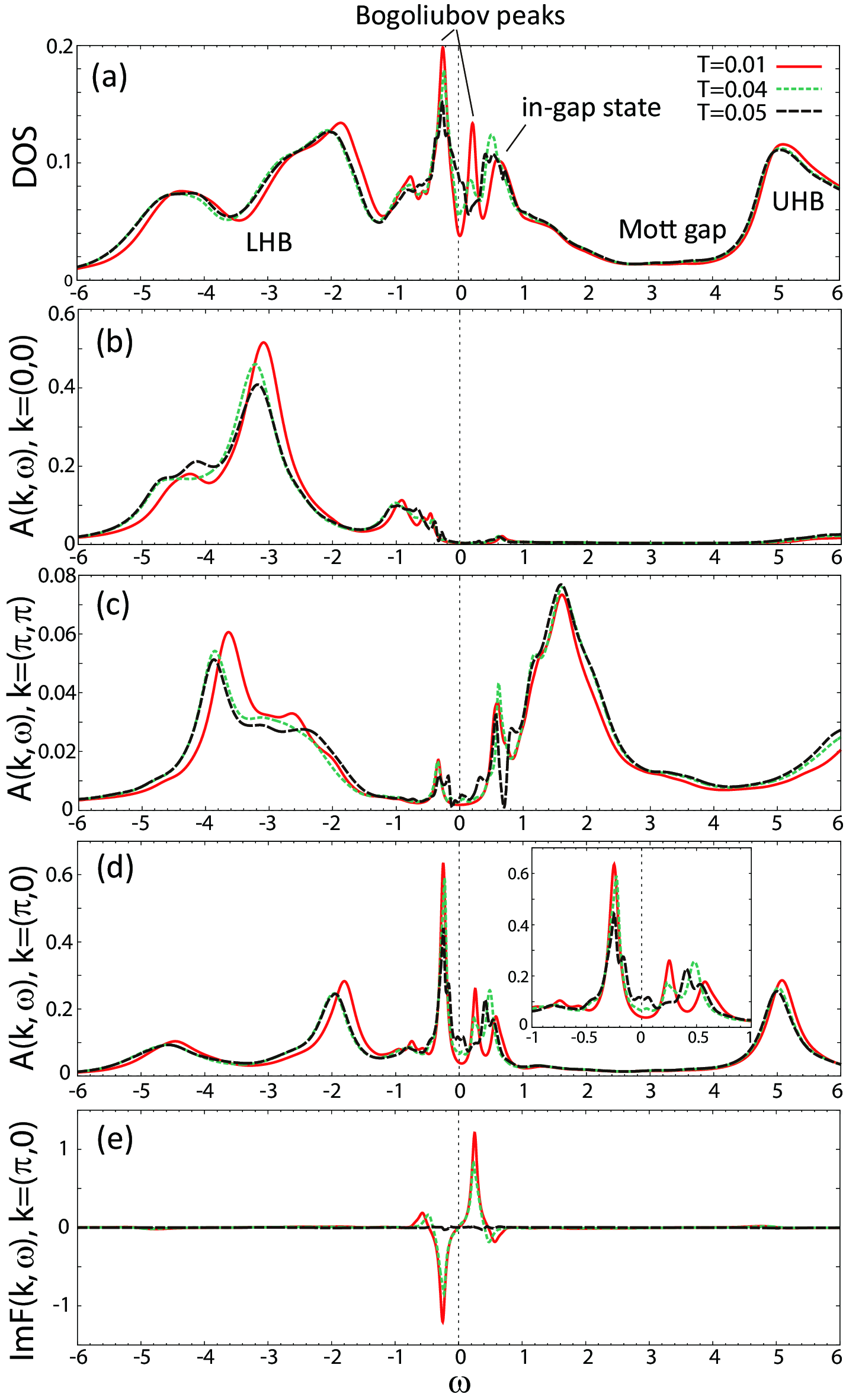}}
\caption{(Color online). (a) Density of states obtained with CDMFT for $n=0.935$ and $U=8$ for several temperatures. 
$\Tc$ is between 0.045 and 0.05.
UHB and LHB denote the upper and lower Hubbard bands, respectively. (b-d) Spectral function at $\Vec{k}=(0,0)$, $(\pi,\pi)$ and $(\pi,0)$. Inset to panel (d) shows the enlarged view of the low-energy part. (e) Im$F$ at $\Vec{k}=(\pi,0)$.}
\label{fig:g_ge_t}
\end{figure}

In this subsection, we discuss the $T$ dependence of the electronic structure for fixed values of $n=0.935$ and $U=8$. We start with discussing the global structure. 
Figure \ref{fig:g_ge_t}(a) shows the local density of states (DOS),
\begin{align}
 \text{DOS}(\w)\equiv -\frac{1}{\pi}\text{Im}G_\text{loc}(\w),
\label{dos}
\end{align}
at several temperatures above and below $\Tc$ ($0.045<\Tc<0.05$). Here $G_\text{loc}$ is the onsite component of the normal Green's function.
As is expected, the high-energy part does not appreciably change with $T$ while the structure for $|\w|<t$ does.
For $2<\w<3$, the DOS is suppressed, which we call the Mott gap, and above the gap the upper Hubbard band (UHB) is located. 
For $T=0.05>\Tc$ (black-dashed curve), we see a suppression of the DOS, particle-hole asymmetric around the Fermi energy ($\w=0$), 
which is identified with the pseudogap. 
Between the pseudogap and the Mott gap, there is a weight around $\w=0.5$, which is called the ingap state.
Below $\Tc$, the Bogoliubov particle-hole symmetric peaks appear around the Fermi energy. Notice that  
the upper Bogoliubov branch appears inside the energy range of the pseudogap above $\Tc$, while 
the lower Bogoliubov branch grows in correspondence of the lower pseudogap edge. Therefore, in the underdoped region the superconducting gap appears competing with the pseudogap for the electronic states above the Fermi level, while it shares with the pseudogap the same electronic states under the Fermi level, as pointed out in Ref. \onlinecite{loret16}.

Figures \ref{fig:g_ge_t}(b)-(d) show the spectral function $A(\Vec{k},\w)$ at the three cluster momenta, $\Vec{k}=(0,0)$, $(\pi,\pi)$ and $(\pi,0)$.
We see that at $\Vec{k}=(0,0)$ and $(\pi,\pi)$ there are no significant low-energy states: Notice that at  $\Vec{k}=(\pi,\pi)$ the overall intensity is pretty small.
It is therefore reasonable to focus on $\Vec{k}=\kAN=(\pi,0)$, where the low-energy structures (i.e., pseudogap, ingap state, Bogoliubov peaks) corresponding to those seen in the DOS are clearly observed.
Figure \ref{fig:g_ge_t}(e) shows the corresponding Im$F$ at $\Vec{k}=\kAN$.
We see that it has a strong intensity only in the low-energy region, $|\w|\lesssim t$.
The peak position of Im$F$ agrees with that of the Bogoliubov peaks in $A(\kAN,\w)$.

\begin{figure}[tb]
\center{\includegraphics[width=0.48\textwidth]{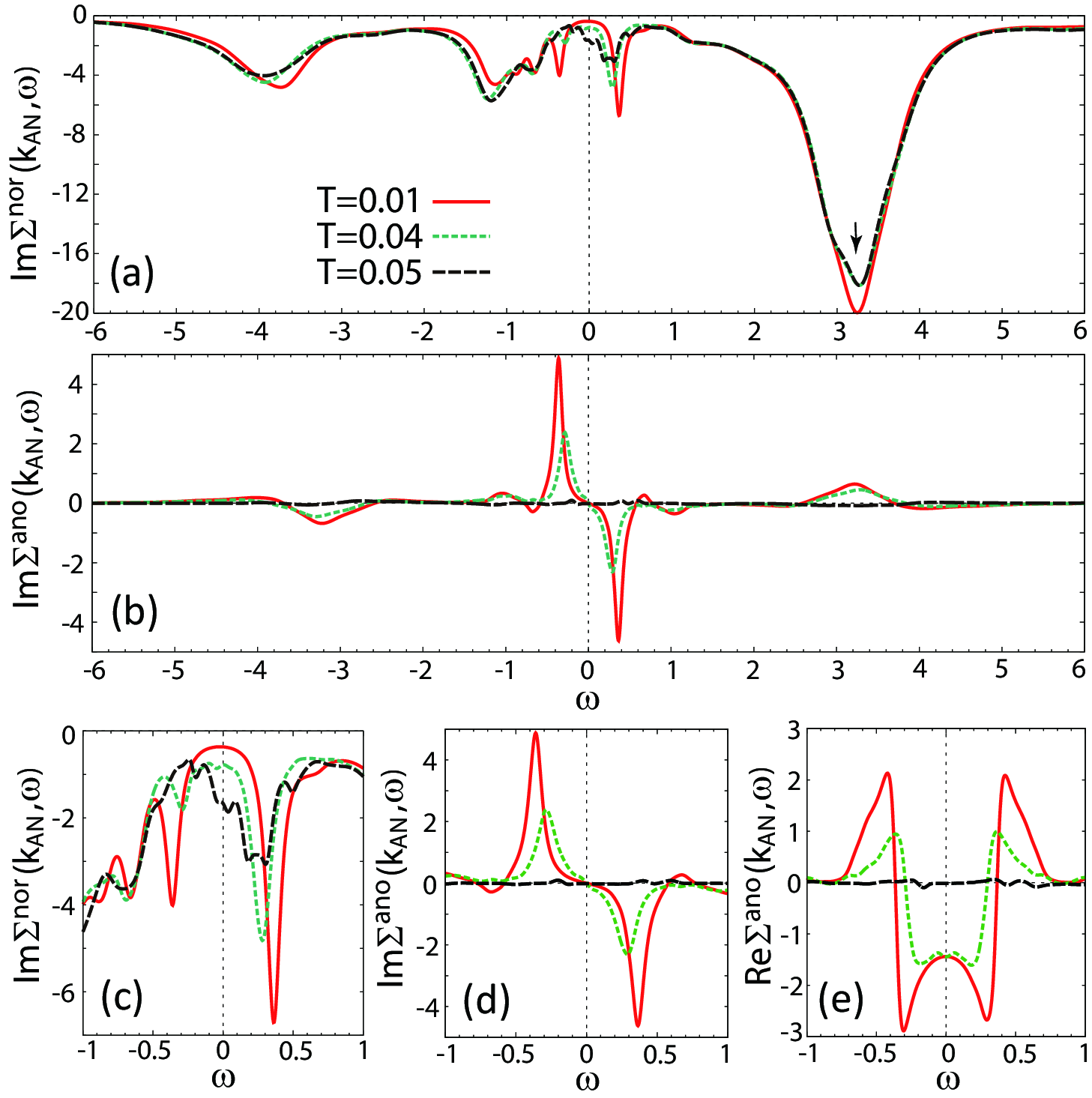}}
\caption{(Color online). (a) Im$\Snor$ and (b) Im$\Sano$ obtained with CDMFT for $n=0.935$ and $U=8$ for several temperatures at $\Vec{k}=\kAN =(\pi,0)$. The black arrow denotes the normal self-energy peak related to the Mott gap. Inset to panel (a) shows the enlarged view of the low-energy part indicated by the grey open box in the main panel.}
\label{fig:sig_ge_t}
\end{figure}

In Fig.~\ref{fig:sig_ge_t} we plot the corresponding self-energy at $\Vec{k}=\kAN$.
In the normal state (black-dashed curve), the peak position of Im$\Snor$ has one-to-one correspondence with the dip position of $A(\kAN,\w)$ in Fig.~\ref{fig:g_ge_t}(d).
The largest self-energy peak at $\w\simeq 3.3$, marked by the arrow, produces the Mott gap in $A(\kAN,\w)$.
Above $\Tc$ the peak of Im$\Snor$ at $\w\simeq 0.2-0.3$ [see the closeup in panel (c)] gives the pseudogap in $A(\kAN,\w)$.
Below $\Tc$, a pair of low-energy peaks appears at electron-hole symmetric positions (at $\w\simeq\pm 0.35$).
The relation between $\Snor$ and $A$ becomes more complicated in the superconducting state due to the appearance of $\Sano$.
Im$\Sano$ in Fig.~\ref{fig:sig_ge_t}(b) or \ref{fig:sig_ge_t}(d) shows strong low-energy peaks at the same positions as Im$\Snor$\cite{sakai16}.
As $T$ decreases, these self-energy peaks acquire more intensity, with slightly shifting to higher energy.

Due to the Kramers-Kr\"onig relation, Re$\Sano$, plotted in Fig.~\ref{fig:sig_ge_t}(e), shows a sign change around the peak positions of Im$\Sano$ and is enhanced around $\w=0$: The low-energy value of Re$\Sano$ is determined by Im$\Sano$ through 
\begin{align}
{\rm Re}\Sano(\kAN,\w=0)=\frac{2}{\pi}\int_0^\infty \frac{d\w'}{\w'}{\rm Im}\Sano(\kAN,\w').
\label{KK}
\end{align}
Hence, the peak energy of Im$\Sano$ characterizes the energy scale below which the superconductivity comes into play.

\begin{figure}[tb]
\center{\includegraphics[width=0.48\textwidth]{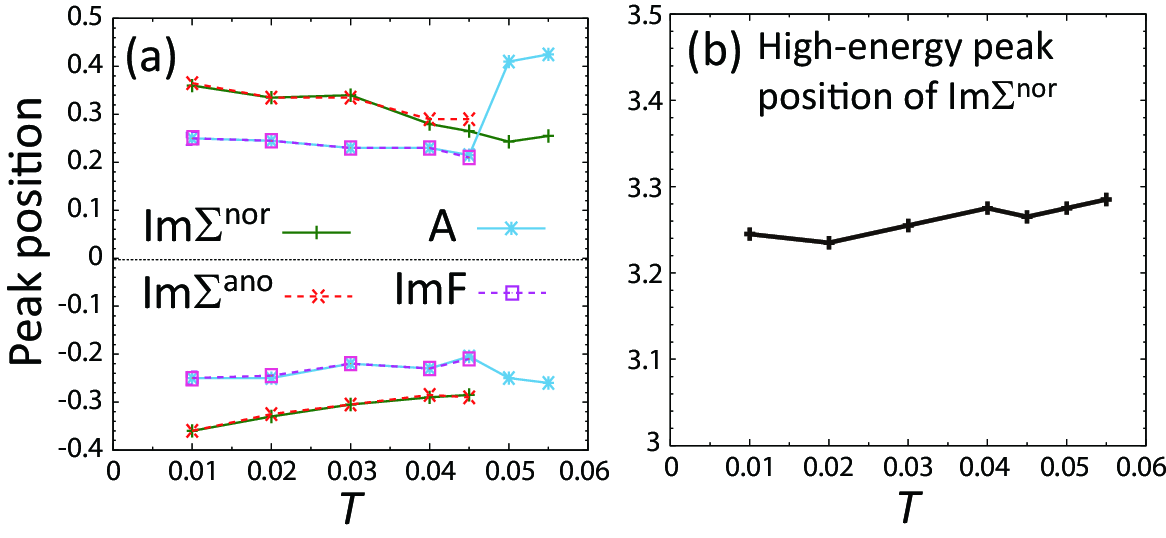}}
\caption{(Color online). $T$ dependence of the peak positions in Im$\Snor$, Im$\Sano$, $A$, and Im$F$ at $\Vec{k}=\kAN$ for $U=8$ and $n=0.935$, where $\Tc$ is just slightly above $0.045$. 
(a) Peak positions related to the low-energy peaks. For Im$\Snor$ for $T=0.05$ [black-dashed curve in Fig.~\ref{fig:sig_ge_t}(a) inset] and $T=0.055$, the low-energy peak shows a small splitting presumably due to an artifact of the finite-size calculation, so that we have taken the averaged value of the splitted peak positions. (b) Position of  the high-energy peak related to the Mott gap. }
\label{fig:peak_t}
\end{figure}

We summarize in Fig.~\ref{fig:peak_t} the $T$ depenence of the peak positions.
For the high-energy peak [panel (b)], the $T$ dependence is just weak. 
More interesting behaviors are found for the low-energy peaks [panel (a)].
First of all, below $\Tc$, the peak positions of the normal and anomalous parts 
match nearly perfectly for both the self-energy and Green's function. 
Above $\Tc$, the anomalous components (and hence the peak positions) vanish, as well as the negative-energy peaks of $\Snor$: The persisting positive-energy branch of $\Snor$ generates the pseudogap and pushes up the 
upper branch of the spectral function $A$ to $\w\sim 0.4$, while the negative-energy branch of $A$ remains fixed at $\w\simeq -0.25$.
This means that the spectral gap [which can be evaluated as the energy difference between the positive- and negative-energy branches of $A$ in Fig.~\ref{fig:peak_t}(a)]
suddenly decreases when $T$ crosses $\Tc$ from above, and this effect is more clearly seen on the unoccupied side of $A$.

Below $\Tc$, the Bogoliubov peaks of Green's function always reside at smaller energy (in absolute value) than those of the self-energy and both tends to decrease as $T$ increases.
The weakly-temperature dependent spectral gap, without closing at $\Tc$, is consistent with $B_{1g}$ (antinodal) Raman experimental result \cite{loret16}, which sees at the first approximation the excitation from an occupied state to unoccupied state.

\subsection{Doping dependence}\label{ssec:akw_ndep}

\begin{figure}[tb]
\center{\includegraphics[width=0.48\textwidth]{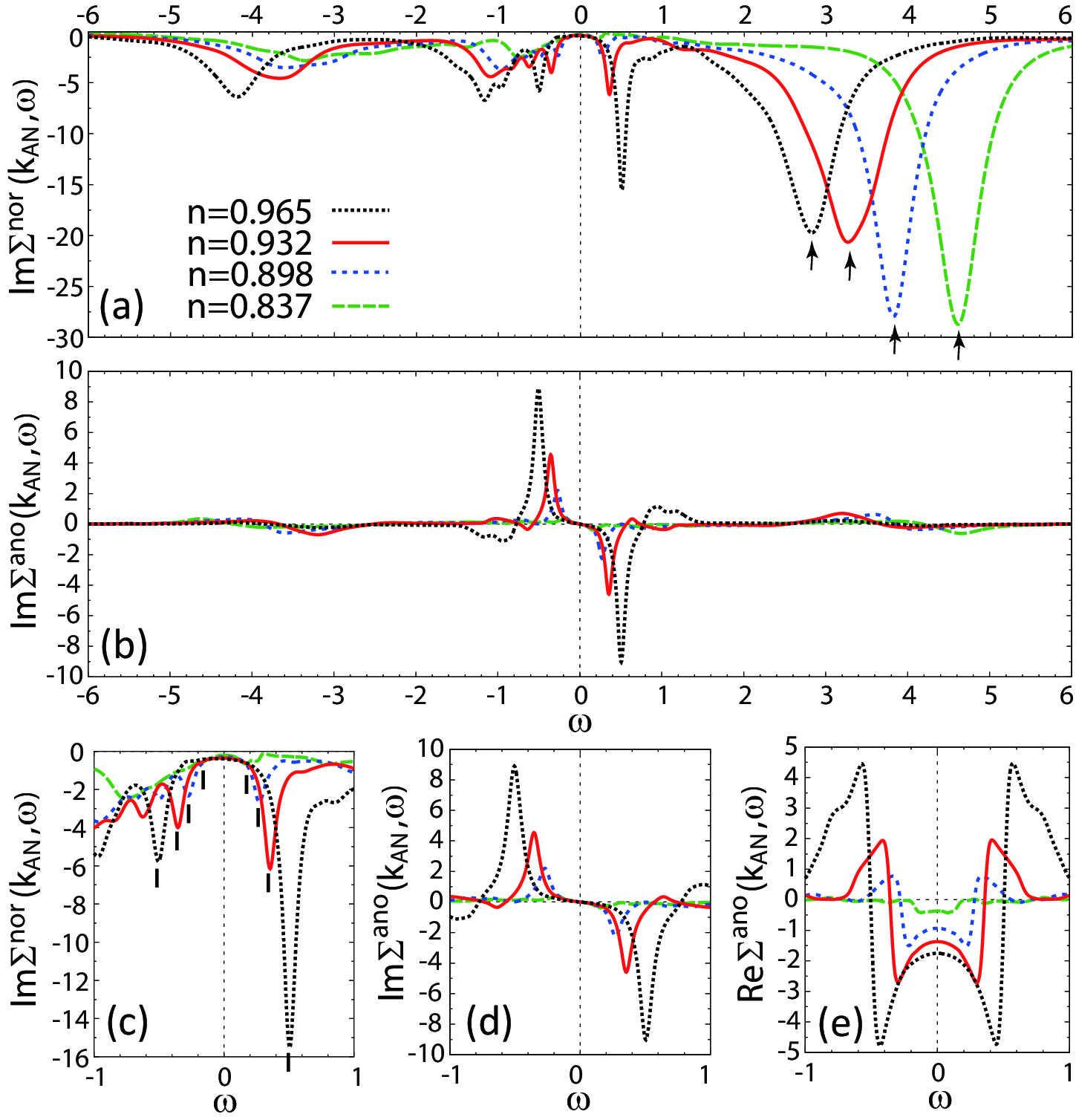}}
\caption{(Color online). Self-energies at $\Vec{k}=\kAN =(\pi,0)$ for various fillings at $T=0.01$ and $U=8$. At all the fillings, the system is in the superconucting state. $n=0.965$ and $0.932$ are in the underdoped region, $n=0.898$ is close to the optimal doping, and $n=0.837$ is in the overdoped region. 
(a) and (b) show the imaginary part of the normal and anomalous components, respectively, in a global energy range. 
(c) and (d) give the enlarged views of (a) and (b), respectively, on the low-energy part. 
(e) shows the real part of the anomalous component.
The black arrows (vertical bars) in panel (a) [(c)] denote the high- (low-)energy peak discussed in the text.}
\label{fig:sig_ge_n}
\end{figure}

In this subsection, we study $n$ dependence of the self-energy structure for fixed $T=0.01$ and $U=8$.
Figure \ref{fig:sig_ge_n} compares the self-energy for several different densities: According to Fig.~\ref{fig:dop}(a), $n=0.965$ and $0.932$ are in the underdoped region,  $n=0.898$ is close to the optimal doping, and $n=0.835$ is in the overdoped region.
As $n$ decreases, the high-energy peak of Im$\Snor$, denoted by arrows in Fig.~\ref{fig:sig_ge_n}(a),  shifts to higher energy, keeping a strong intensity.
On the other hand, the low-energy peak, denoted by vertical bars in Fig.~\ref{fig:sig_ge_n}(c), loses its intensity and shifts to lower energy with doping.
With further doping, the low-energy peak merges with other structures arising on the higher-energy side of the peak, becoming almost undiscernible.

Im$\Sano$ in Fig.~\ref{fig:sig_ge_n}(b) shows only small intensity at high energies ($\w\sim3 - 5$) while it shows prominent low-energy peaks at $\w\sim0.2-0.5$ in agreement with previous results \cite{haule07,maier08,kyung09,civelli09PRL,gull13PRL,gull15,sakai16}.
As the doping increases, the low-energy peaks shift to lower energy and get weakened [see panel (d)], in accordance with the behavior of the peaks in Im$\Snor$.
In Fig.~\ref{fig:sig_ge_n}(e), we see that the low-energy peaks of Im$\Sano$ enhances Re$\Sano$ at low $\w$. This effect is stronger for smaller doping.
How $\Sano$ vanishes at zero doping is an intriguing but delicate issue, which we will address elsewhere.

\begin{figure}[tb]
\center{\includegraphics[width=0.48\textwidth]{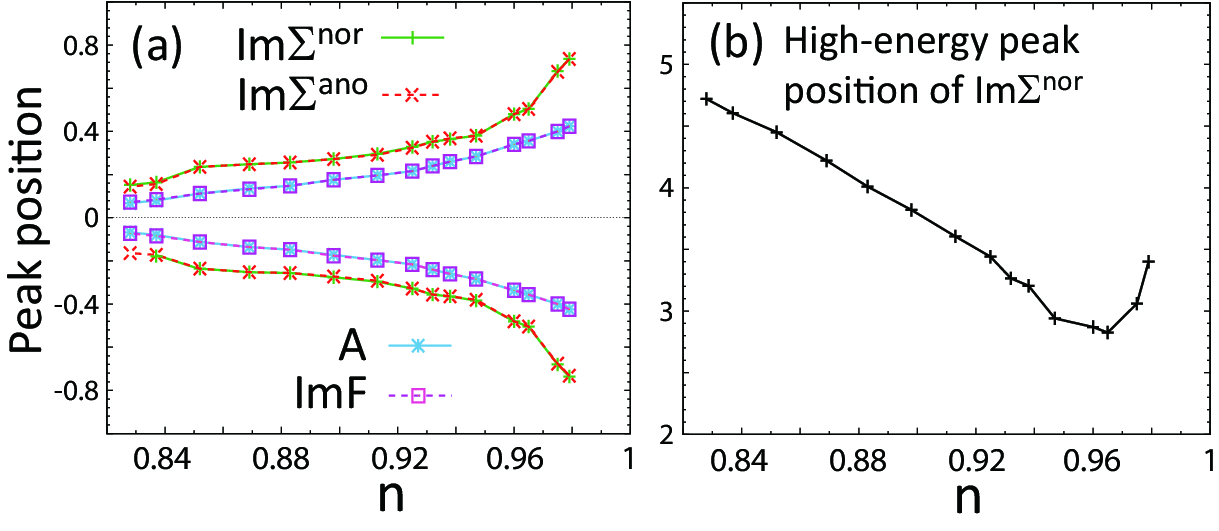}}
\caption{(Color online). Doping dependence of the peak positions of Im$\Snor$, Im$\Sano$, $A$, and Im$F$ at $\Vec{k}=\kAN$ for $U=8$ and $T=0.01$. All the data are in the superconducting state. 
(a) Low-energy peak positions. For $n<0.83$, the negative-energy peak of Im$\Snor$ is not well resolved. (b) High-energy peak of Im$\Snor$ related to the Mott gap. 
}
\label{fig:peak_n}
\end{figure}

We summarize in Fig.~\ref{fig:peak_n} the density depenence of the positions of these peaks.
All the data are in the superconducting state so that the peaks are located at electron-hole symmetric positions. The correspondence between the peak positions of normal and anomalous components also holds. Notice that for $n<0.83$ the negative-energy peak of Im$\Snor$ merges with other structures and is not well resolved. 
Panel (a) shows that the absolute value of the energy position of the low-energy peak decreases with doping and those of $A$ and Im$F$ approach zero, indicating the closure of the superconducting gap at $n$ smaller than 0.82. 
The peak positions of the self-energy are always higher in absolute value than those of Green's function.
On the other hand, panel (b) plots the high-energy peak position, marked by arrows in Fig.~\ref{fig:sig_ge_n}(a),  
which decreases as $n$ increases up to $\sim 0.96$, and then turns up in the vicinity of the Mott transition at half filling.
This behavior can be understood by a combination of the behavior of the atomic-limit self-energy and a screening effect due to the doped holes, as we shall elaborate in Sec.~\ref{sec:discuss}. 

\subsection{$U$ dependence}\label{ssec:akw_udep}

\begin{figure}[tb]
\center{\includegraphics[width=0.48\textwidth]{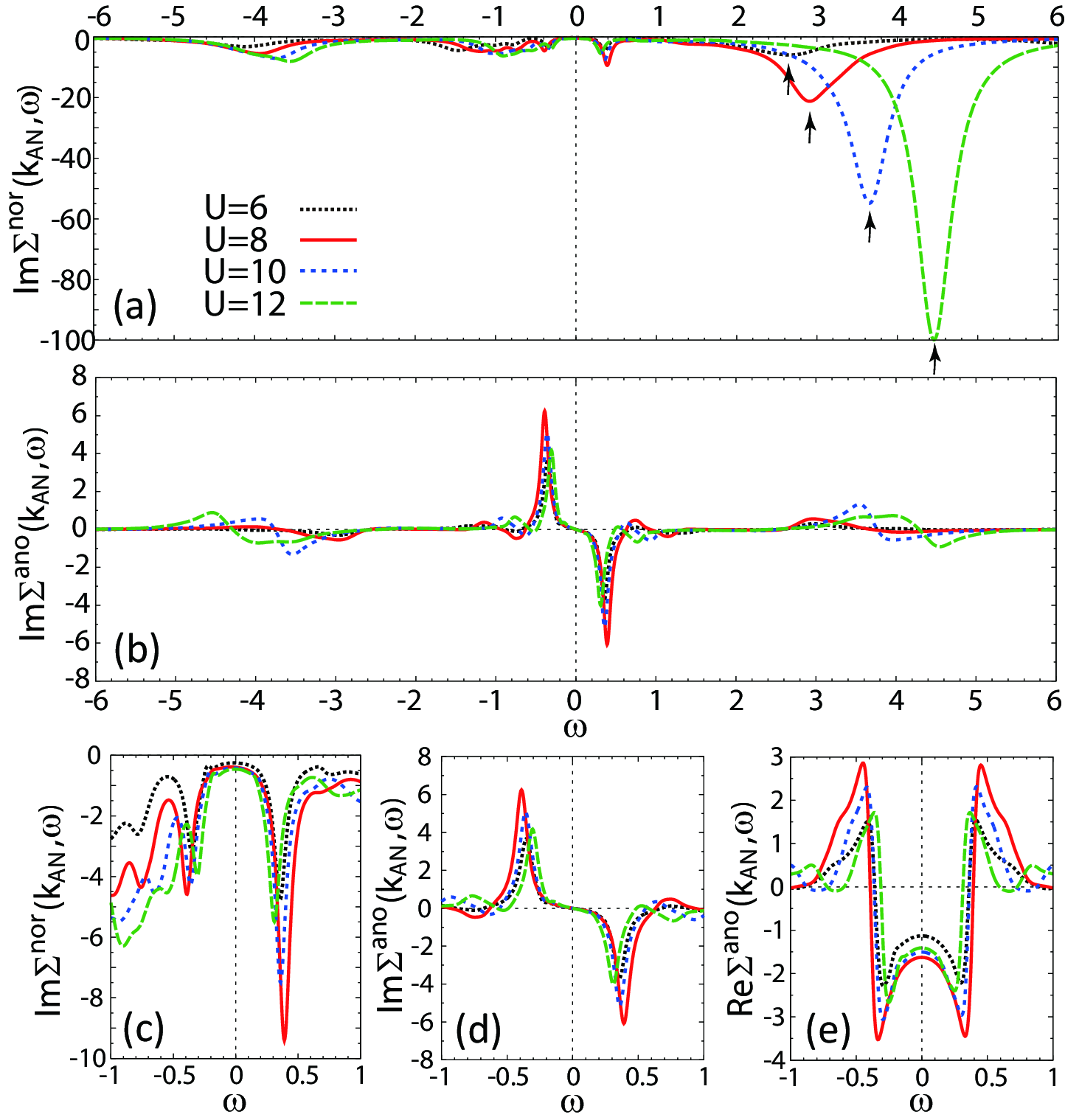}}
\caption{(Color online). Self-energies at $\Vec{k}=\kAN =(\pi,0)$ for several values of $U$ at $T=0.01$ and $n=0.95$. For all the values of $U$, the system is in the superconducting state. 
(a) and (b) show the imaginary part of the normal and anomalous components, respectively, in a global energy range. The black arrow denotes the normal self-energy peak related to the Mott gap.
(c) and (d) give the enlarged views of (a) and (b), respectively, on the low-energy part. (e) shows the real part of the anomalous component.
}
\label{fig:sig_ge_u}
\end{figure}

In this subsection, we study the $U$ dependence of the self-energy structure for fixed $n=0.95$ and $T=0.01$.
Figure \ref{fig:sig_ge_u} plots the CDMFT self-energies at $\Vec{k}=\kAN$ for $U=6$, 8, 10 and 12.
For all these values of $U$, the system is in the superconducting state. 
As $U$ increases, the high-energy peak in Im$\Snor$, denoted by black arrow in panel (a), acquires more intensity and shifts to higher energy. 
On the other hand, the low-energy peaks reside always around $\w=\pm 0.3-0.4$ and does not significantly depend on $U$ [panels (c) and (d)].
The peak intensity also rather weakly depends on $U$ while it is maximized around $U=8$ and is suppressed for larger $U$.
Accordingly, the enhancement of Re$\Sano$ at low energy is also maximized around $U=8$ [panel (e)].

\begin{figure}[tb]
\center{\includegraphics[width=0.48\textwidth]{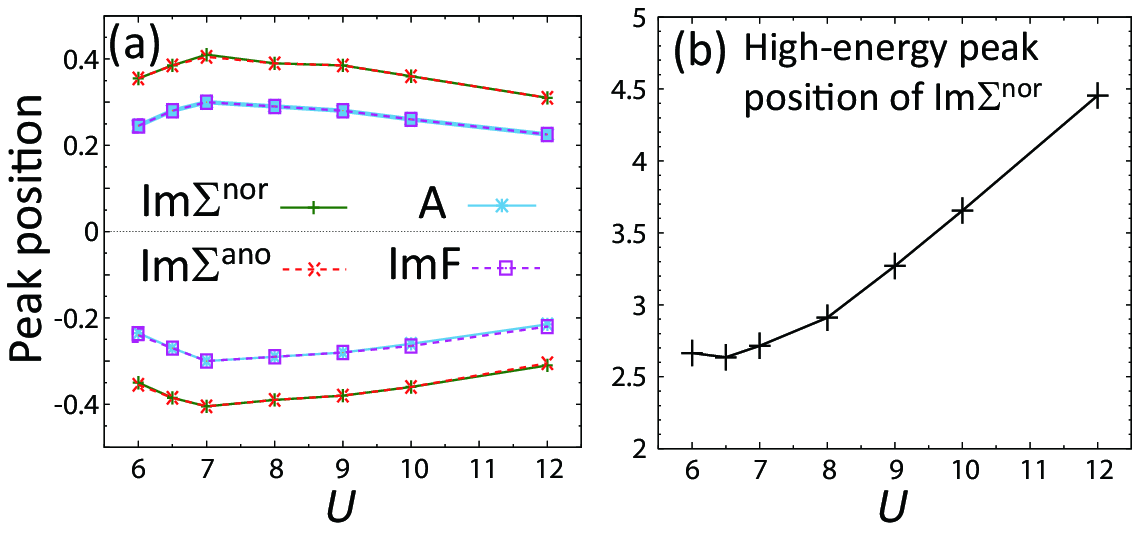}}
\caption{(Color online). $U$ dependence of the peak positions Im$\Snor$, Im$\Sano$, $A$, and Im$F$ at $\Vec{k}=\kAN$ for $T=0.01$ and $n=0.95$. 
(a) Low-energy peaks. (b) High-energy peak of Im$\Snor$ related to the Mott gap. }
\label{fig:peak_u}
\end{figure}

We summarize the $U$ dependence of the peak positions in Fig.~\ref{fig:peak_u}.
Both for the self-energies and Green's functions, the energies of the low-energy peaks depend rather weakly on $U$ and show a maximum (in absolute value) around $U=7$ [panel (a)].
On the other hand, the position of the high-energy peak is lifted by $U$, almost linearly for $U>8$, as is expected for the behavior in the atomic limit (see Sec.~\ref{ssec:fit_udep}).

\section{Interpretation by hidden fermion}\label{sec:interpret}
In this section, we explore the low-energy self-energy structure, obtained with the CDMFT, in terms of the phenomenology described in Sec.~\ref{ssec:tcfm}.

\subsection{Fitting by hidden-fermion model}
\begin{figure}[tb]
\center{\includegraphics[width=0.48\textwidth]{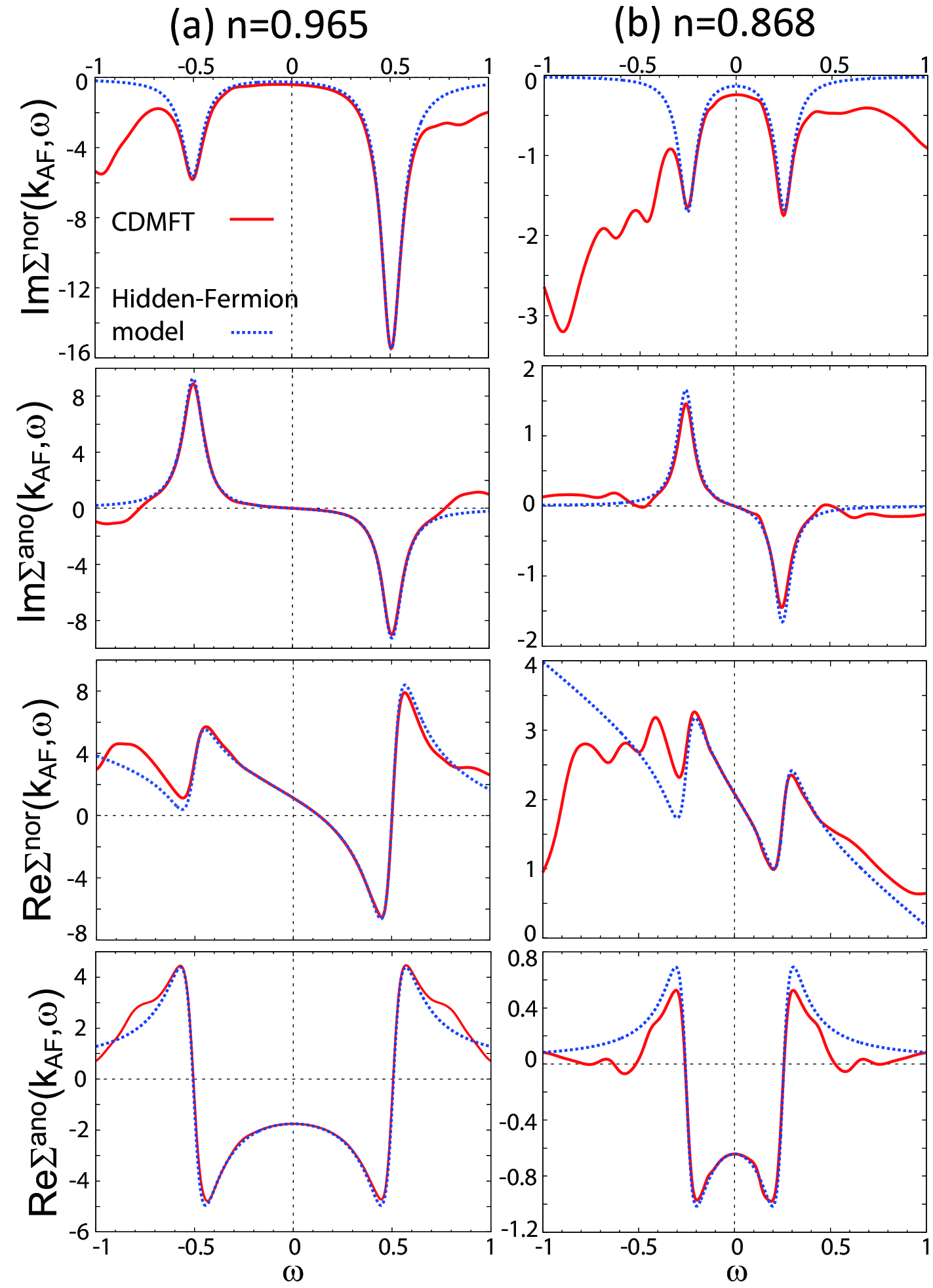}}
\caption{(Color online). (a) Low-energy part of the self-energy calculated with the CDMFT (red-solid curve) for $U=8$, $T=0.01$ and $n=0.965$, and that fitted with Eq.~(\ref{sigc_sc1he}) (blue-dotted curve) of the hidden-fermion model. (b) The same for $n=0.868$. }
\label{fig:sig_fit}
\end{figure}

We fit the CDMFT self-energy with Eq.~(\ref{sigc_sc1he}), i.e., 
\begin{align*}
 \Snor(\Vec{k},\w)&\simeq \frac{V_1(\Vec{k})^2(\w+\e_{f_1}(\Vec{k}))}{\w^2-\e_{f_1}(\Vec{k})^2-D_{f_1}(\Vec{k})^2} 
 + \tilde{a}(\Vec{k})-\bk\w,\nonumber\\
 \Sano(\Vec{k},\w)&\simeq \frac{-V_1(\Vec{k})^2 D_{f_1}(\Vec{k})}{\w^2-\e_{f_1}(\Vec{k})^2-D_{f_1}(\Vec{k})^2}+\tilde{D}_{\rm HE}(\Vec{k}),
 \tag{\ref{sigc_sc1he}}
\end{align*}
where we use the same energy-broadening factor $\h(\w)$ as that used to calculate the CDMFT self-energy.
Here we focus on the low-energy region including the self-energy peaks, and determine the parameters in Eq.~(\ref{sigc_sc1he}) by a least-square fitting, according to the procedure described in Sec.~\ref{sssec:fit}.

In Fig.~\ref{fig:sig_fit}, we show to what extent the fitting works, for two selected dopings, $n=0.965$ and 0.868: The former is in the underdoped region, and the latter is in the slightly overdoped region. 
Between these two dopings, the low-energy peaks of the self-energy are well separated from other higher-energy structures and hence the fitting works well. 
On the other hand, for $n>0.965$ and $n<0.868$, other structures come into the relevant low-energy range, as discussed in Sec.~\ref{ssec:akw_ndep}, so that the fitting does not work so well. 
The fitting parameters are determined below in the region where the low-energy peaks are well resolved.

\subsection{Doping dependence of fitting parameters}\label{ssec:fit_ndep}

\begin{figure}[tb]
\center{\includegraphics[width=0.48\textwidth]{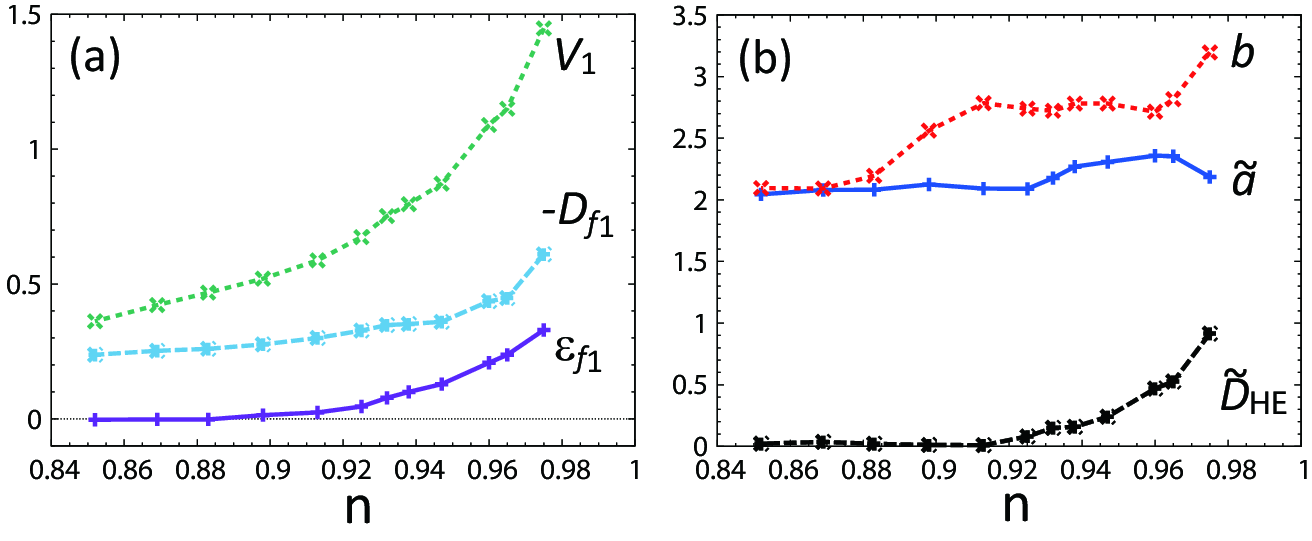}}
\caption{(Color online). $n$ dependence of the fitting parameters for $U=8$ and $T=0.01$ at $\Vec{k}=\kAN$. 
(a) Parameters related to the low-energy pole $f_1$. (b) Parameters related to the corrections from high-energy poles.}
\label{fig:para_n}
\end{figure}

Figure \ref{fig:para_n} plots the parameters, obtained with the procedure described above, against the electron density $n$ for fixed $U=8$ and $T=0.01$. 
We see that all of $\ef$, $|\Df|$ and $\Vf$ decrease as $n$ decreases
\footnote{Note that the sign of $\Df$ depends on which of ($\pi,0$) and (0,$\pi$) we choose to plot for a $d$-wave symmetry-broken state and does not matter.}.
The decrease of $\Vf$ reflects the decreasing intensity of the self-energy poles while $\Vf$ is still substantial in the overdoped region ($n<0.88$).
Since $\ef$ and $|\Df|$, as well as $\tilde{D}_{\rm HE}$, are rather flat in the overdoped region, the decrease of $\Vf$ is responsible for the decrease of the order parameter there.

Interestingly, $\ef$ is virtually zero in the optimal to overdoped region.
In fact, a smaller value of $\ef$ is helpful for enhancing the superconductivity because, as Eq.~(\ref{KK}) shows, a pole at a lower energy can enhance the superconducting gap more \footnote{This is true even when we consider the superconducting gap function $\D$ rather than $\Sano$ because $\D$ in the hidden-fermion model has a pole at $\w=\pm\sqrt{\ef^2+\Df^2+\Vf^2}$ \cite{sakai16}}.
Since $\ef$ is the source of the electron-hole asymmetry in the low-energy part of $\Snor$ [and eventually of $A(\kAN,\w)$] as is evident from Eq.~(\ref{sigc_sc1he}), the disappearance of $\ef$ manifests the presence of the electron-hole symmetry in this region, in accordance with previous CDMFT results \cite{kyung06PRB1,sordi12PRL}, as well as with STM experimental results \cite{pushp09}.

Figure \ref{fig:para_n}(b) shows the parameters related to the corrections due to high-energy poles.
$|\Df|\gg|\DHE|$ holds for $n<0.97$, suggesting that $f_1$ indeed enhances the superconductivity considerably.  
In the vicinity of $n=1$, $\DHE$, as well as $b$, increases because the self-energy structure other than $f_1$ comes into the low-energy region, which eventually breaks down the fitting for $n>0.98$.  

\subsection{Temperature dependence of fitting parameters}\label{ssec:fit_tdep}

\begin{figure}[tb]
\center{\includegraphics[width=0.48\textwidth]{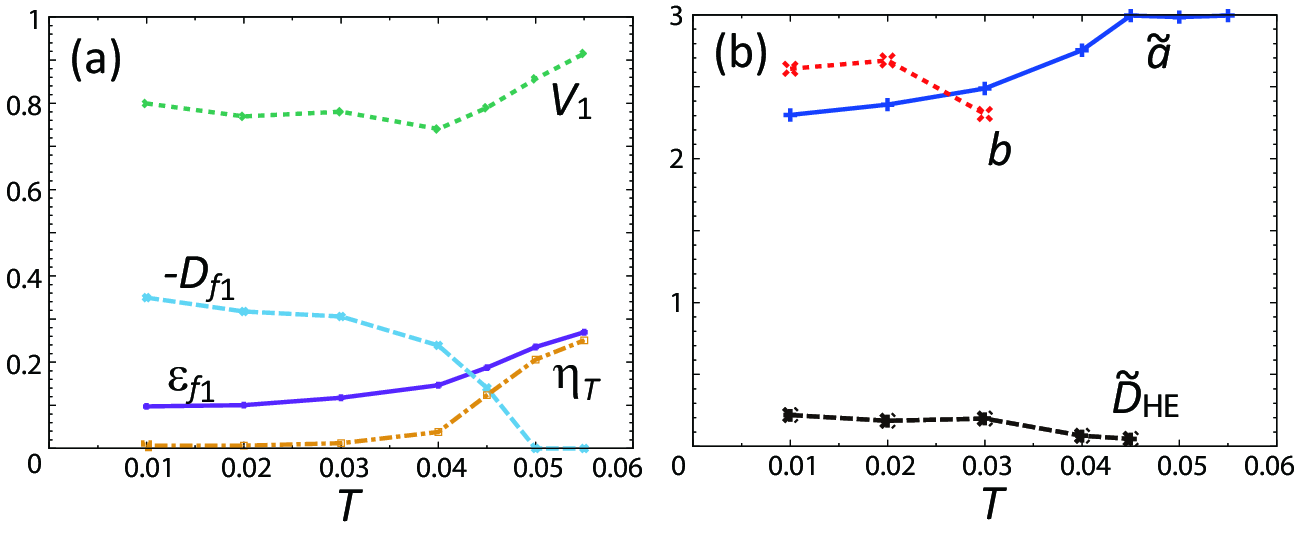}}
\caption{(Color online). $T$ dependence of the fitting parameters for $U=8$ and $n=0.935$ at $\Vec{k}=\kAN$. 
(a) Parameters related to the low-energy pole $f_1$. (b) Parameters related to the corrections from high-energy poles.
For $T\geq 0.04$, a numerical fluctuation of $\Snor$ at low-energy, which is seen in Fig.~\ref{fig:sig_ge_t}(c) and is likely ascribed to a finite size effect, prevents a reliable evaluation of $b$ so that it is not plotted there.
$T_{\rm c}$ is between 0.045 and 0.05, as seen in Fig.~\ref{fig:dop}(b). }
\label{fig:para_t}
\end{figure}

In order to take account of thermal fluctuations approximately, we introduce a temperature-dependent broadening factor $\h_T$ as another fitting parameter in Eq.~(\ref{sigc_sc1he}): We replace $\w$ with $\w+i[\h(\w)+\h_T]$.
Figure \ref{fig:para_t} plots the fitting parameters obtained for the CDMFT self-energies at $\Vec{k}=\kAN$, $U=8$ and $n=0.935$.

We see a continuous change of the fitting parameters from the superconducting to normal states.
This indicates that the same hidden fermion $f_1$ is relevant for both states \cite{sakai16}.
Below $\Tc\simeq 0.045-0.05$, we see that i) $\Vf$ does not appreciably depend on $T$ (and also the high-energy contribution $\tilde{a}$ varies smoothly).
On the other hand, $|\Df|$ increases rapidly just below $\Tc$ and then saturates (as the high-energy contribution $\DHE$ does). 
$\ef$ decreases with decreasing $T$. 
$\h_T$ increases with increasing $T$ as is expected.
The small $\h_T$ (which is virtually zero for $T<0.03$) justifies neglecting this factor for the fitting of the $T=0.01$ results, as we have done in other sections.

\subsection{$U$ dependence of fitting parameters}\label{ssec:fit_udep}

\begin{figure}[tb]
\center{\includegraphics[width=0.48\textwidth]{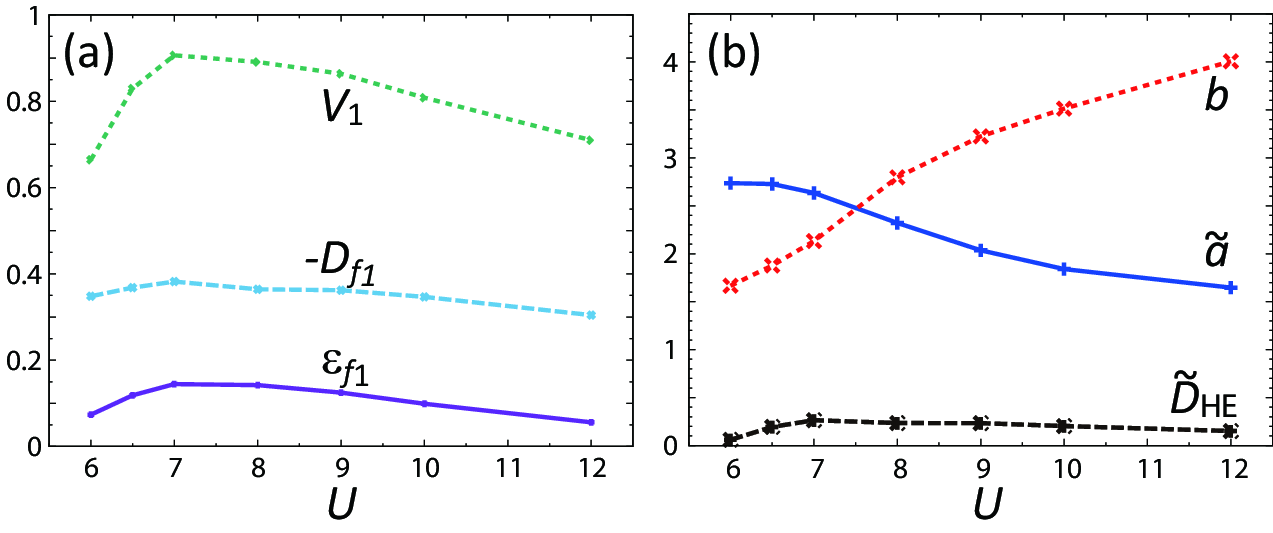}}
\caption{(Color online). $U$ dependence of the fitting parameters for $T=0.01$ and $n=0.95$ at $\Vec{k}=\kAN$.
(a) Parameters related to the low-energy pole $f_1$. (b) Parameters related to the corrections from high-energy poles. }
\label{fig:para_u}
\end{figure}

Figure \ref{fig:para_u} shows $U$ dependence of the parameters for $T=0.01$ and $n=0.95$ at $\Vec{k}=\kAN$.
As mentioned in Sec.~\ref{ssec:dop}, we have restricted our study to the strongly-correlated region, $U>6$.
We find that $\Vf$, $|\Df|$ and $\ef$ take maxima around $U=7$:
The rapid decrease of $\Vf$ for the smaller $U$ suggests a strong-coupling origin of $f_1$ while it slowly decreases for the larger $U$, too.
The behavior of $\ef$ indicates that the electron-hole asymmetry is maximized around $U=7$.
The increase of $b$ in panel (b) suggests that the renormalization due to the high-energy poles becomes more significant for larger $U$, despite that the high-energy pole shifts to a higher energy with $U$ [see Fig.~\ref{fig:peak_u}(b)]. This is because the intensity of the high-energy pole increases more rapidly ($\propto U^2$), as we shall discuss in Sec.~\ref{sec:discuss}.

\section{DISCUSSIONS}\label{sec:discuss}
 
In the atomic limit at zero temperature, the electron self-energy is given by
\begin{align}
\Snor_{\rm atom}(\w)=\frac{n}{2}U+\frac{\frac{n}{2}(1-\frac{n}{2})U^2}{\w-(1-\frac{n}{2})U},
\label{atom}
\end{align}
which has a pole, at $\varepsilon_{\rm atom}=(1-\frac{n}{2})U$, of an intensity $V_{\rm atom}^2=\frac{n}{2}(1-\frac{n}{2})U^2$.
This pole was identified with the pole appearing in the normal and superconducting states of the strongly-attactive Hubbard model \cite{sakai15}.
The present results for the repulsive Hubbard model, however, show that the behavior of $\ef$ and $\Vf$ is different from the above $\varepsilon_{\rm atom}$ and $V_{\rm atom}$.  
For instance, $\ef$ and $\Vf$ monotonically increase with $n$ [Fig.~\ref{fig:para_n}(a)]
while $\varepsilon_{\rm atom}$ decreases and $V_{\rm atom}$ is nonmonotonic with $n$;
$\ef$ and $\Vf$ are nonmonotonic with $U$ [Fig. \ref{fig:para_u}(a)] while 
$\varepsilon_{\rm atom}$ and $V_{\rm atom}$ monotonically increase with $U$.

Equation (\ref{atom}) is rather consistent with the behavior of the high-energy pole related to the Mott gap:
A roughly linear decrease with $n$ for $n<0.96$ [Fig.~\ref{fig:peak_n}(b)], and a roughly linear increase of the peak position [Fig.~\ref{fig:peak_u}(b)], as well as a rapid increase of the peak intensity [Fig.~\ref{fig:sig_ge_u}(a)], with $U$ are all qualitatively consistent with Eq.~(\ref{atom}). However, quantitatively, the peak position is somewhat lower than the value expected from $(1-\frac{n}{2})U$. This could be an artifact of the finite bath in the ED while it is more likely that a screening due to other electrons and holes (which are not present in the atomic limit) reduces the effective onsite interaction. 
The rapid reduction of the screening is consistent with the quick rise of the peak position 
observed in the vicinity of $n=1$, where the reduced screening further suppresses the 
appearance of the doublon-hole excitations.

On the other hand, the origin of the low-energy pole $f_1$ is more involved. 
As the nonmonotonic $U$ dependence in Fig.~\ref{fig:para_u}(a) suggests, it emerges from a correlation effect
and is most stabilized in the nonperturbative regime where $U\sim$ bandwidth.
From the $n$ dependence plotted in Fig.~\ref{fig:para_n}(a), the relevance of the Mott physics is also apparent.
The $T$ dependence plotted in Fig.~\ref{fig:para_t}(a) also indicates that the $f_1$ fermion is the same as that generating the pseudogap above $\Tc$.

A key observation would be that for all the parameters studied, $\ef$ is positive (or very tiny in the overdoped region). 
The hybridization between $c$ and $f_1$ generates the pseudogap above $\Tc$, and at the same time spectral weights above and below the pseudogap. This spectral weight below the pseudogap describes the quasiparticle while the weight above the pseudogap gives the ingap state, generated by doping the Mott insulator.
Namely, the origin of $f_1$ should be attributed to the spectral-weight transfer due to the doping to the Mott insulator. The transfer yields a low-energy degree of freedom ($f_1$) in addition to $c$,  and then they hybridize with each other to open a gap (pseudogap). Even below $\Tc$, they keep hybridizing, and it enhances the superconductivity through a consequent pole in $\Sano$.

An immediate problem is therefore to clarify the nature of the ingap state. A hint, extracted from Fig.~\ref{fig:para_n}(a), is that $\ef$ decreases with doping in the underdoped region: This behavior is qualitatively opposite to that of electron and quasiparticle since $\e_c(\Vec{k})$ [Eq.~(\ref{disp})] increases with doping (due to the decrease of $\mu$),
suggesting that the holes are created dominantly as the $f_1$ excitation rather than $c$ excitation when the doping concentration is decreased. 
The relevance of the hole excitation to $f_1$ is consistent with the momentum dependence of $\ef$, too \cite{sakai16}.
One interesting possibility consistent with these observations is a low-energy excitonic bound state of an electron and a hole, proposed in Ref.~\onlinecite{yamaji11PRL}.  

In the low-energy effective model consisting only of the $c$ and $f_1$ degrees of freedom,
$c$ can be replaced by a ``quasipartice'' $\tilde{c}$ in which the original electron is renormalized by $f_\a$'s other than $f_1$. This renormalization within our 2$\times$2 CDMFT calculation is represented by $b$ plotted in Figs.~\ref{fig:para_n}(b), \ref{fig:para_t}(b) and \ref{fig:para_u}(b).
$\tilde{c}$ is a ``bright'' fermion because it gives a major contrubution to the single-particle spectra and is visible.
On the other hand, $f_1$ is a ``dark'' fermion because it gives zero to Green's function and is not directly visible in the single-particle spectra.
The electron transits between the bistable bright and dark states, as expressed by the hybridization. 
The bright and dark fermions may be identified with fractionalized degrees of freedom of 
the original electron.
The opposite doping dependences of  $\ef$ and $\e_c$ may suggest the change in the relative stability and dominance of the dark and bright excitations (induced by a level crossing) leading naturally to a double-well structure in total energy as a function of doping.  This may have a connection to a first-order transition discussed in Ref.~\onlinecite{misawa14} and in references therein (including cluster-DMFT works by Refs.~\onlinecite{capone06,aichhorn07,sordi12PRL}). Detailed understanding on this matter is an intriguing future issue.

\section{SUMMARY}\label{sec:summary}
We have discussed the frequency-dependent structure of the electron self-energy in the pseudogap and superconducting states of the 2D Hubbard model, based on the numerical results calculated within the 2$\times$2 CDMFT.
The finite-temperature extension of the ED, which we used to solve the CDMFT impurity problem, allowed us to study properties on the real-frequency axis precisely. 
We have systematically investigated how the low-energy real-frequency structure of the self-energy at the antinodal point changes with temperature, electron density, and the Hubbard $U$.

In the normal state above $\Tc$, Im$\Snor$ shows a peak at low energy, which produces the pseudogap in the spectral function. Another prominent peak in Im$\Snor$ is located at $\w\sim \frac{U}{2}$, which produces the Mott gap.
The latter peak does not change appreciably as the temperature is lowered below $\Tc$.
From its dependences on $n$ and $U$, we conclude that this peak originates from the property
in the atomic limit.

On the other hand, the former low-energy peak shows more nontrivial dependences on the parameters.
In the superconducting state, this peak appears also in Im$\Sano$, doubled at particle-hole symmetric energy positions.
Since the peak in Im$\Sano$ enhances the low-energy value of Re$\Sano$ (i.e., strengthens the superconductivity) according to Eq.~(\ref{KK}), it is essential for revealing the high-$\Tc$ superconducting mechanism.
We have therefore performed a detailed analysis of this peak.

In our previous work \cite{sakai16}, we showed that the nontrivial property of this peak (i.e., peak-to-peak cancellation between normal and anomalous contributions to Green's function) is consistent with the self-energy peak
 generated by a hybridization of the electron with a hidden fermionic excitation, and demonstrated a quantitative fitting at one selected point of ($T$, $n$, $U$). 
In the present study, we have carried out the fitting for various points in the ($T$, $n$, $U$) space and determined the hidden-fermion parameters, $\ef$, $\Vf$, and $\Df$.
The $T$, $n$, and $U$ dependences of these parameters characterize the hidden fermion $f_1$.

The continuous $T$-dependent change of the hidden-fermion parameters across $\Tc$ (Fig.~\ref{fig:para_t}) confirms our previous argument \cite{sakai16} that the same hidden fermion $f_1$ is playing a role in the pseudogap and superconductivity.
The increase of $\Vf$ and $|\Df|$ close to half filling  (Fig.~\ref{fig:para_n}) implies the relevance of the Mott physics to $f_1$.
The increase of $\ef$ with $n$ further implies the increasing relevance of $f_1$ in a hole excitation.
The nonmonotonic $U$ dependence of the hidden-fermion parameters (Fig.~\ref{fig:para_u}), peaked at $U\sim$ bandwidth, suggests a nonperturbative nature of $f_1$.
These numerical results impose strong constraints on possible candidates of the hidden fermion,
and eventually the mechanism of the high-$\Tc$ superconductivity.

Interestingly, the hidden-fermion energy $\ef$ goes to zero [Fig.~\ref{fig:para_n}(a)] in correspondence with the optimal doping.
Owing to Eq.~(\ref{KK}), in fact, closer $\ef$ is to the Fermi level, higher is the 
hidden-fermion peak contribution to superconductivity. 
As we further increase doping, however, while $\ef$ stays essentially zero, 
$V_1$ decreases [Fig.~\ref{fig:para_n}(a)], reducing in this way the intensity of the hidden-fermion peak, and consequently its contribution to the superconductivity. 
Hence for doping greater than the optimal one, we expect $\Tc$ decreasing. The appearance of the hidden fermionic excitation $f_1$
close to the Mott insulator can explain then at the same time the pseudogap in the normal state and why superconductivity has a high $\Tc$
displaying a domelike shape as a function of doping.

As a final remark, we expect that the simple expression of the self-energy, Eq.~(\ref{sigc_sc1he}), 
will be useful in analysing one-electron spectroscopic experimental data (e.g., scanning-tunneling and angle-resolved photoemission spectroscopies) of general strongly-correlated superconductors \footnote{A caution is needed to discuss two or more electron processes within the hidden-fermion Hamiltonian because the vertex corrections in the original many-body system are not straightforwardly taken into account in the hidden-fermion Hamiltonian.}.
In particular for the cuprates, the dependence of the parameters on $T$, $n$, and $U$ shown in Figs.~\ref{fig:para_t}, \ref{fig:para_n}, and  \ref{fig:para_u}, respectively, will also be helpful.
In fact, in the normal state, Eq.~(\ref{signor_ns1he}) can be regarded as a generalization 
of the Yang-Rice-Zhang ansatz \cite{yang06}, which has been useful in analysing experimental results for the pseudogap state of the cuprates \cite{rice12}, and Eq.~(\ref{sigc_sc1he}) is the superconducting extension of Eq.~(\ref{signor_ns1he}).
Furthermore, the hidden-fermion representation of the self-energy, which replaces the 
many-body effect with the one-body Hamiltonian having additional fermionic degrees of freedom,
 may offer interesting applications, as recently exemplified in Ref.~\onlinecite{seki16} for a combination of the tetrahedron method and the cluster perturbation theory \cite{senechal00}. 

\section*{Acknowledgment}
S.S. thanks A. Liebsch for fruitful discussions in developing the numerical simulation code used in the present study.
S.S. is supported by JSPS KAKENHI (Grant No. 26800179).
M.I. is supported by the Computational Materials Science
Initiative (CMSI), HPCI Strategic Programs for
Innovative Research (SPIRE), and RIKEN Advanced
Institute for Computational Science (AICS) through
HPCI System Research Project (Grants No. hp120043,
No. hp120283, No. hp130007, No. hp140215, and
No. hp150211), from MEXT, Japan.

\section*{APPENDIX A: Derivation of Eqs.~(\ref{signor_ns}) and (\ref{sigc_sc})}
We derive Eq.~(\ref{sigc_sc}) by integrating out the $f$ degrees of freedom in the model (\ref{tcf_sc}).
Equation (\ref{signor_ns}) can then be obtained straightforwardly by substituting $D_c=D_{f_\a}=0$ in Eq.~(\ref{sigc_sc}).
Since the contribution of each $f_\a$ to the $c$-fermion self-energy is additive, it is sufficient to derive the equation for a single $f$ fermion.
Therefore, we start with the following action, 
\begin{align}
&S[c^\dagger,c,f^\dagger,f]\nonumber\\
=\ &\sum_{\Vec{k},n>0}\{-i\w_n(c_{\Vec{k}\ua,n}^\dagger c_{\Vec{k}\ua,n}+c_{-\Vec{k}\da,-n} c_{-\Vec{k}\da,-n}^\dagger \nonumber\\
+\ &f_{\Vec{k}\ua,n}^\dagger f_{\Vec{k}\ua,n}+ f_{-\Vec{k}\da,-n} f_{-\Vec{k}\da,-n}^\dagger)\nonumber\\
+\ &H_{\Vec{k},n}[c^\dagger,c,f^\dagger,f]\} \label{action}
\end{align}
with
\begin{align}
&H_{\Vec{k},n}[c^\dagger,c,f^\dagger,f]\nonumber\\
\equiv\ &
\e_c(\Vec{k})(c_{\Vec{k}\ua,n}^\dagger c_{\Vec{k}\ua,n}+c_{-\Vec{k}\da,-n}^\dagger c_{-\Vec{k}\da,-n})\nonumber\\
+\ &\e_f(\Vec{k})(f_{\Vec{k}\ua,n}^\dagger f_{\Vec{k}\ua,n}+f_{-\Vec{k}\da,-n}^\dagger f_{-\Vec{k}\da,-n})\nonumber\\
+\ &V(\Vec{k})(c_{\Vec{k}\ua,n}^\dagger f_{\Vec{k}\ua,n}+c_{-\Vec{k}\da,-n}^\dagger f_{-\Vec{k}\da,-n}+\text{H. c.})\nonumber\\
-\ &D_c(\Vec{k})(c_{\Vec{k}\ua,n} c_{-\Vec{k}\da,-n}+\text{H. c.})\nonumber\\
-\ &D_f(\Vec{k})(f_{\Vec{k}\ua,n} f_{-\Vec{k}\da,-n}+\text{H. c.}).
\end{align}
Here, the subscript $n$ in $c^\dagger$, $c$, $f^\dagger$ and $f$ represents each Fourier component at the Matsubara frequency $\w_n=(2n-1)\pi T$. In the following, we sometimes abbreviate $\Vec{k}$ in the argument of $V$, $\e_{c,f}$ and $D_{c,f}$ just for the sake of brevity. However, we can always recover it since the Hamiltonian is diagonal in $\{\Vec{k},-\Vec{k}\}$.

With the action, the partition function is given by
\begin{align}
Z=\int \mathcal{D}c^\dagger \mathcal{D}c \mathcal{D}f^\dagger \mathcal{D}f \exp\left(-S[c^\dagger,c,f^\dagger,f]\right).
\label{z}
\end{align}
We define the effective action $S^\text{eff}$ for the $c$ fermion by integrating out the $f$ degrees of freedom in Eq.~(\ref{z}), i.e.,
\begin{align}
&\frac{1}{Z^\text{eff}}\exp\left(-S^\text{eff}[c^\dagger,c]\right)\nonumber\\
=\ &\frac{1}{Z}
\int \mathcal{D}f^\dagger \mathcal{D}f \exp\left(-S[c^\dagger,c,f^\dagger,f]\right)
\label{Seff}
\end{align}
with a constant $Z^\text{eff}\equiv \int \mathcal{D}c^\dagger \mathcal{D}c \exp\left(-S^\text{eff}[c^\dagger,c]\right)$.

The $(\Vec{k},n)$ component in Eq.~(\ref{action}) is recast into
\begin{align}
&(E_\Vec{k}-i\w_n)\tilde{f}_{\Vec{k}\ua,n}^\dagger \tilde{f}_{\Vec{k}\ua,n} 
+ (E_\Vec{k}+i\w_n) \tilde{f}_{-\Vec{k}\da,-n}^\dagger \tilde{f}_{-\Vec{k}\da,-n} \nonumber\\
&-\frac{V^2(i\w_n+\e_f)}{E_\Vec{k}^2+\w_n^2}c_{\Vec{k}\ua,n}^\dagger c_{\Vec{k}\ua,n}\nonumber\\
&+\frac{V^2(i\w_n-\e_f)}{E_\Vec{k}^2+\w_n^2}c_{-\Vec{k}\da,-n}^\dagger c_{-\Vec{k}\da,-n}\nonumber\\
&-\frac{V^2 D_f}{E_\Vec{k}^2+\w_n^2}(c_{\Vec{k}\ua,n} c_{-\Vec{k}\da,-n}+ \text{H.c.}),
\label{action2}
\end{align}
where $E_\Vec{k}\equiv\sqrt{\e_f(\Vec{k})^2+D_f(\Vec{k})^2}$ and 
\begin{align}
\tilde{f}_{\Vec{k}\ua,n}^\dagger&\equiv u_\Vec{k} f_{\Vec{k}\ua,n}^\dagger+v_\Vec{k} f_{-\Vec{k}\da,-n}\nonumber\\
&+\frac{V}{E_\Vec{k}-i\w_n}(u_\Vec{k} c_{\Vec{k}\ua,n}^\dagger+v_\Vec{k} c_{-\Vec{k}\da,-n}),\nonumber\\
\tilde{f}_{\Vec{k}\ua,n}&\equiv u_\Vec{k} f_{\Vec{k}\ua,n}+v_\Vec{k} f_{-\Vec{k}\da,-n}^\dagger
\nonumber\\
&+\frac{V}{E_\Vec{k}-i\w_n}(u_\Vec{k} c_{\Vec{k}\ua,n}+v_\Vec{k} c_{-\Vec{k}\da,-n}^\dagger),\nonumber\\
\tilde{f}_{-\Vec{k}\da,-n}^\dagger&\equiv u_\Vec{k} f_{-\Vec{k}\da,-n}^\dagger-v_\Vec{k} f_{\Vec{k}\ua,n}\nonumber\\
&-\frac{V}{E_\Vec{k}+i\w_n}(u_\Vec{k} c_{-\Vec{k}\da,-n}^\dagger-v_\Vec{k} c_{\Vec{k}\ua,n}),\nonumber\\
\tilde{f}_{-\Vec{k}\da,-n}&\equiv u_\Vec{k} f_{-\Vec{k}\da,-n}-v_\Vec{k} f_{\Vec{k}\ua,n}^\dagger\nonumber\\
&-\frac{V}{E_\Vec{k}+i\w_n}(u_\Vec{k} c_{-\Vec{k}\da,-n}-v_\Vec{k} c_{\Vec{k}\ua,n}^\dagger)
\end{align}
with $u_\Vec{k}\equiv\sqrt{\left(1+\frac{\e_f(\Vec{k})}{E_\Vec{k}}\right)/2}$ and $v_\Vec{k}\equiv\sqrt{\left(1-\frac{\e_f(\Vec{k})}{E_\Vec{k}}\right)/2}$.

Since the integration of the $\tilde{f}^\dagger$ and $\tilde{f}$ degrees of freedom in Eq.~(\ref{action2}) gives just a constant, we obtain
\begin{align}
S^\text{eff}[c^\dagger,c]
&=\sum_{\Vec{k},n>0}\left\{\left[\e_c-\frac{V^2(i\w_n+\e_f)}{E_\Vec{k}^2+\w_n^2}\right]
    c_{\Vec{k}\ua,n}^\dagger c_{\Vec{k}\ua,n}\right.\nonumber\\
&+\left[\e_c+\frac{V^2(i\w_n-\e_f)}{E_\Vec{k}^2+\w_n^2}\right]
   c_{-\Vec{k}\da,-n}^\dagger c_{-\Vec{k}\da,-n}\nonumber\\
&\left.-\left[D_c+\frac{V^2 D_f}{E_\Vec{k}^2+\w_n^2}\right](c_{\Vec{k}\ua,n} c_{-\Vec{k}\da,-n}+ \text{H. c.})\right\}
\label{action3}
\end{align}
This equation shows that the self-energy correction for the $c$ fermions is described by
\begin{align}
&\Snor(\Vec{k},i\w_n)=-\frac{V^2(i\w_n+\e_f)}{E_\Vec{k}^2+\w_n^2},\nonumber\\
&\Sano(\Vec{k},i\w_n)= D_c+\frac{V^2 D_f}{E_\Vec{k}^2+\w_n^2}.
\label{sigc_iw}
\end{align}

As one can see from the fact that the terms proportional to $V^2$ depend only on the $f$-fermion parameters, these terms are additive when multiple $f$ fermions are introduced.
Then, by replacing $i\w_n$ with $\w+i\eta$ in these terms, we obtain Eq.~(\ref{sigc_sc})
\footnote{The $\w$-independent term $s(\Vec{k})$ in Eq.~(\ref{sigc_sc}) should be included in the dispersion term of $c$, as is expressed in the first term of Eq.~(\ref{tcf_ns}).}.
For the normal state, substituting $D_c=D_f=0$ in the above equations, we obtain Eq.~(\ref{signor_ns}).
Another derivation of Eq.~(\ref{signor_ns}) is found in Ref.~\onlinecite{seki16}.

\section*{APPENDIX B: Derivation of the sum rule Eq.~(\ref{sumrule_sc})}
Following Ref.~\onlinecite{seki11}, we calculate the second moment,
\begin{align}
M_{\Vec{k}}^2=\oint \frac{d\w}{2\pi i}\w^2 G(\Vec{k},\w),
\label{moment}
\end{align}
of the normal Green's function, which is written in the superconducting state as
\begin{align}
&G(\Vec{k},\w)=\nonumber\\
&\left[ \w-\e_c(\Vec{k})-\Snor(\Vec{k},\w)
-\frac{\Sano(\Vec{k},\w)^2}{\w+\e_c(\Vec{k})+\Snor(\Vec{k},-\w)^\ast} \right]^{-1}.
\end{align}
With the high-frequency expansion of Eq.~(\ref{sigc_sc}), we obtain 
\begin{align}
\Snor(\Vec{k},\w)&=s(\Vec{k})+\frac{1}{\w}\sum_\a V_\a(\Vec{k})^2 + O(\w^{-2}),\nonumber\\
\Sano(\Vec{k},\w)&=D_c(\Vec{k})-\frac{1}{\w^2}\sum_\a V_\a(\Vec{k})^2 D_{f_\a}(\Vec{k}) + O(\w^{-4}),
\end{align}
and hence
\begin{align}
G(\Vec{k},\w)&=\frac{1}{\w}+\frac{1}{w^2}(\e_c(\Vec{k})+s(\Vec{k}))\nonumber\\
&+\frac{1}{\w^3}\left\{\sum_\a V_\a(\Vec{k})^2 +(\e_c(\Vec{k})+s(\Vec{k}))^2+D_c(\Vec{k})^2\right\}\nonumber\\
&+ O(\w^{-4}).
\end{align}
Substituting this equation into Eq.~(\ref{moment}) leads to
\begin{align}
M_{\Vec{k}}^2=\sum_\a V_\a(\Vec{k})^2 +(\e_c(\Vec{k})+s(\Vec{k}))^2+D_c(\Vec{k})^2.
\label{moment2}
\end{align}

Meanwhile, $M_{\Vec{k}}^2$ can alternatively be calculated from the commutation relation of $c_{\Vec{k}\s}$ and $c^\dagger_{\Vec{k}\s}$ with the Hamiltonian (\ref{hubbard}) \cite{harris67}, which results in 
\begin{align}
M_{\Vec{k}}^2=\e_c(\Vec{k})^2+\e_c(\Vec{k})nU+\frac{n}{2}U^2.
\end{align}
Comparing this equantion with Eq.~(\ref{moment2}), and using Eq.~(\ref{sk}),
we obtain the sum rule, Eq.~(\ref{sumrule_sc}).

\bibliography{ref}

\begin{thebibliography}{61}%
\makeatletter
\providecommand \@ifxundefined [1]{%
 \@ifx{#1\undefined}
}%
\providecommand \@ifnum [1]{%
 \ifnum #1\expandafter \@firstoftwo
 \else \expandafter \@secondoftwo
 \fi
}%
\providecommand \@ifx [1]{%
 \ifx #1\expandafter \@firstoftwo
 \else \expandafter \@secondoftwo
 \fi
}%
\providecommand \natexlab [1]{#1}%
\providecommand \enquote  [1]{``#1''}%
\providecommand \bibnamefont  [1]{#1}%
\providecommand \bibfnamefont [1]{#1}%
\providecommand \citenamefont [1]{#1}%
\providecommand \href@noop [0]{\@secondoftwo}%
\providecommand \href [0]{\begingroup \@sanitize@url \@href}%
\providecommand \@href[1]{\@@startlink{#1}\@@href}%
\providecommand \@@href[1]{\endgroup#1\@@endlink}%
\providecommand \@sanitize@url [0]{\catcode `\\12\catcode `\$12\catcode
  `\&12\catcode `\#12\catcode `\^12\catcode `\_12\catcode `\%12\relax}%
\providecommand \@@startlink[1]{}%
\providecommand \@@endlink[0]{}%
\providecommand \url  [0]{\begingroup\@sanitize@url \@url }%
\providecommand \@url [1]{\endgroup\@href {#1}{\urlprefix }}%
\providecommand \urlprefix  [0]{URL }%
\providecommand \Eprint [0]{\href }%
\providecommand \doibase [0]{http://dx.doi.org/}%
\providecommand \selectlanguage [0]{\@gobble}%
\providecommand \bibinfo  [0]{\@secondoftwo}%
\providecommand \bibfield  [0]{\@secondoftwo}%
\providecommand \translation [1]{[#1]}%
\providecommand \BibitemOpen [0]{}%
\providecommand \bibitemStop [0]{}%
\providecommand \bibitemNoStop [0]{.\EOS\space}%
\providecommand \EOS [0]{\spacefactor3000\relax}%
\providecommand \BibitemShut  [1]{\csname bibitem#1\endcsname}%
\let\auto@bib@innerbib\@empty
\bibitem [{\citenamefont {Metzner}\ and\ \citenamefont
  {Vollhardt}(1989)}]{metzner89}%
  \BibitemOpen
  \bibfield  {author} {\bibinfo {author} {\bibfnamefont {W.}~\bibnamefont
  {Metzner}}\ and\ \bibinfo {author} {\bibfnamefont {D.}~\bibnamefont
  {Vollhardt}},\ }\href {\doibase 10.1103/PhysRevLett.62.324} {\bibfield
  {journal} {\bibinfo  {journal} {Phys. Rev. Lett.}\ }\textbf {\bibinfo
  {volume} {62}},\ \bibinfo {pages} {324} (\bibinfo {year} {1989})}\BibitemShut
  {NoStop}%
\bibitem [{\citenamefont {Georges}\ \emph {et~al.}(1996)\citenamefont
  {Georges}, \citenamefont {Kotliar}, \citenamefont {Krauth},\ and\
  \citenamefont {Rozenberg}}]{georges96}%
  \BibitemOpen
  \bibfield  {author} {\bibinfo {author} {\bibfnamefont {A.}~\bibnamefont
  {Georges}}, \bibinfo {author} {\bibfnamefont {G.}~\bibnamefont {Kotliar}},
  \bibinfo {author} {\bibfnamefont {W.}~\bibnamefont {Krauth}}, \ and\ \bibinfo
  {author} {\bibfnamefont {M.~J.}\ \bibnamefont {Rozenberg}},\ }\href {\doibase
  10.1103/RevModPhys.68.13} {\bibfield  {journal} {\bibinfo  {journal} {Rev.
  Mod. Phys.}\ }\textbf {\bibinfo {volume} {68}},\ \bibinfo {pages} {13}
  (\bibinfo {year} {1996})}\BibitemShut {NoStop}%
\bibitem [{\citenamefont {Kotliar}\ \emph {et~al.}(2001)\citenamefont
  {Kotliar}, \citenamefont {Savrasov}, \citenamefont {P\'alsson},\ and\
  \citenamefont {Biroli}}]{kotliar01}%
  \BibitemOpen
  \bibfield  {author} {\bibinfo {author} {\bibfnamefont {G.}~\bibnamefont
  {Kotliar}}, \bibinfo {author} {\bibfnamefont {S.~Y.}\ \bibnamefont
  {Savrasov}}, \bibinfo {author} {\bibfnamefont {G.}~\bibnamefont {P\'alsson}},
  \ and\ \bibinfo {author} {\bibfnamefont {G.}~\bibnamefont {Biroli}},\ }\href
  {\doibase 10.1103/PhysRevLett.87.186401} {\bibfield  {journal} {\bibinfo
  {journal} {Phys. Rev. Lett.}\ }\textbf {\bibinfo {volume} {87}},\ \bibinfo
  {pages} {186401} (\bibinfo {year} {2001})}\BibitemShut {NoStop}%
\bibitem [{\citenamefont {Maier}\ \emph
  {et~al.}(2005{\natexlab{a}})\citenamefont {Maier}, \citenamefont {Jarrell},
  \citenamefont {Pruschke},\ and\ \citenamefont {Hettler}}]{maier05RMP}%
  \BibitemOpen
  \bibfield  {author} {\bibinfo {author} {\bibfnamefont {T.}~\bibnamefont
  {Maier}}, \bibinfo {author} {\bibfnamefont {M.}~\bibnamefont {Jarrell}},
  \bibinfo {author} {\bibfnamefont {T.}~\bibnamefont {Pruschke}}, \ and\
  \bibinfo {author} {\bibfnamefont {M.~H.}\ \bibnamefont {Hettler}},\ }\href
  {\doibase 10.1103/RevModPhys.77.1027} {\bibfield  {journal} {\bibinfo
  {journal} {Rev. Mod. Phys.}\ }\textbf {\bibinfo {volume} {77}},\ \bibinfo
  {pages} {1027} (\bibinfo {year} {2005}{\natexlab{a}})}\BibitemShut {NoStop}%
\bibitem [{\citenamefont {S\'en\'echal}\ \emph {et~al.}(2000)\citenamefont
  {S\'en\'echal}, \citenamefont {Perez},\ and\ \citenamefont
  {Pioro-Ladri\`ere}}]{senechal00}%
  \BibitemOpen
  \bibfield  {author} {\bibinfo {author} {\bibfnamefont {D.}~\bibnamefont
  {S\'en\'echal}}, \bibinfo {author} {\bibfnamefont {D.}~\bibnamefont {Perez}},
  \ and\ \bibinfo {author} {\bibfnamefont {M.}~\bibnamefont
  {Pioro-Ladri\`ere}},\ }\href {\doibase 10.1103/PhysRevLett.84.522} {\bibfield
   {journal} {\bibinfo  {journal} {Phys. Rev. Lett.}\ }\textbf {\bibinfo
  {volume} {84}},\ \bibinfo {pages} {522} (\bibinfo {year} {2000})}\BibitemShut
  {NoStop}%
\bibitem [{\citenamefont {Potthoff}\ \emph {et~al.}(2003)\citenamefont
  {Potthoff}, \citenamefont {Aichhorn},\ and\ \citenamefont
  {Dahnken}}]{potthoff03}%
  \BibitemOpen
  \bibfield  {author} {\bibinfo {author} {\bibfnamefont {M.}~\bibnamefont
  {Potthoff}}, \bibinfo {author} {\bibfnamefont {M.}~\bibnamefont {Aichhorn}},
  \ and\ \bibinfo {author} {\bibfnamefont {C.}~\bibnamefont {Dahnken}},\ }\href
  {\doibase 10.1103/PhysRevLett.91.206402} {\bibfield  {journal} {\bibinfo
  {journal} {Phys. Rev. Lett.}\ }\textbf {\bibinfo {volume} {91}},\ \bibinfo
  {pages} {206402} (\bibinfo {year} {2003})}\BibitemShut {NoStop}%
\bibitem [{\citenamefont {Lichtenstein}\ and\ \citenamefont
  {Katsnelson}(2000)}]{lichtenstein00}%
  \BibitemOpen
  \bibfield  {author} {\bibinfo {author} {\bibfnamefont {A.~I.}\ \bibnamefont
  {Lichtenstein}}\ and\ \bibinfo {author} {\bibfnamefont {M.~I.}\ \bibnamefont
  {Katsnelson}},\ }\href {\doibase 10.1103/PhysRevB.62.R9283} {\bibfield
  {journal} {\bibinfo  {journal} {Phys. Rev. B}\ }\textbf {\bibinfo {volume}
  {62}},\ \bibinfo {pages} {R9283} (\bibinfo {year} {2000})}\BibitemShut
  {NoStop}%
\bibitem [{\citenamefont {Maier}\ \emph
  {et~al.}(2005{\natexlab{b}})\citenamefont {Maier}, \citenamefont {Jarrell},
  \citenamefont {Schulthess}, \citenamefont {Kent},\ and\ \citenamefont
  {White}}]{maier05PRL}%
  \BibitemOpen
  \bibfield  {author} {\bibinfo {author} {\bibfnamefont {T.~A.}\ \bibnamefont
  {Maier}}, \bibinfo {author} {\bibfnamefont {M.}~\bibnamefont {Jarrell}},
  \bibinfo {author} {\bibfnamefont {T.~C.}\ \bibnamefont {Schulthess}},
  \bibinfo {author} {\bibfnamefont {P.~R.~C.}\ \bibnamefont {Kent}}, \ and\
  \bibinfo {author} {\bibfnamefont {J.~B.}\ \bibnamefont {White}},\ }\href
  {\doibase 10.1103/PhysRevLett.95.237001} {\bibfield  {journal} {\bibinfo
  {journal} {Phys. Rev. Lett.}\ }\textbf {\bibinfo {volume} {95}},\ \bibinfo
  {pages} {237001} (\bibinfo {year} {2005}{\natexlab{b}})}\BibitemShut
  {NoStop}%
\bibitem [{\citenamefont {Capone}\ and\ \citenamefont
  {Kotliar}(2006)}]{capone06}%
  \BibitemOpen
  \bibfield  {author} {\bibinfo {author} {\bibfnamefont {M.}~\bibnamefont
  {Capone}}\ and\ \bibinfo {author} {\bibfnamefont {G.}~\bibnamefont
  {Kotliar}},\ }\href {\doibase 10.1103/PhysRevB.74.054513} {\bibfield
  {journal} {\bibinfo  {journal} {Phys. Rev. B}\ }\textbf {\bibinfo {volume}
  {74}},\ \bibinfo {pages} {054513} (\bibinfo {year} {2006})}\BibitemShut
  {NoStop}%
\bibitem [{\citenamefont {Aichhorn}\ \emph {et~al.}(2007)\citenamefont
  {Aichhorn}, \citenamefont {Arrigoni}, \citenamefont {Huang},\ and\
  \citenamefont {Hanke}}]{aichhorn07}%
  \BibitemOpen
  \bibfield  {author} {\bibinfo {author} {\bibfnamefont {M.}~\bibnamefont
  {Aichhorn}}, \bibinfo {author} {\bibfnamefont {E.}~\bibnamefont {Arrigoni}},
  \bibinfo {author} {\bibfnamefont {Z.~B.}\ \bibnamefont {Huang}}, \ and\
  \bibinfo {author} {\bibfnamefont {W.}~\bibnamefont {Hanke}},\ }\href@noop {}
  {\bibfield  {journal} {\bibinfo  {journal} {Phys. Rev. Lett.}\ }\textbf
  {\bibinfo {volume} {99}},\ \bibinfo {pages} {257002} (\bibinfo {year}
  {2007})}\BibitemShut {NoStop}%
\bibitem [{\citenamefont {Haule}\ and\ \citenamefont
  {Kotliar}(2007)}]{haule07}%
  \BibitemOpen
  \bibfield  {author} {\bibinfo {author} {\bibfnamefont {K.}~\bibnamefont
  {Haule}}\ and\ \bibinfo {author} {\bibfnamefont {G.}~\bibnamefont
  {Kotliar}},\ }\href {\doibase 10.1103/PhysRevB.76.104509} {\bibfield
  {journal} {\bibinfo  {journal} {Phys. Rev. B}\ }\textbf {\bibinfo {volume}
  {76}},\ \bibinfo {pages} {104509} (\bibinfo {year} {2007})}\BibitemShut
  {NoStop}%
\bibitem [{\citenamefont {Kancharla}\ \emph {et~al.}(2008)\citenamefont
  {Kancharla}, \citenamefont {Kyung}, \citenamefont {S\'en\'echal},
  \citenamefont {Civelli}, \citenamefont {Capone}, \citenamefont {Kotliar},\
  and\ \citenamefont {Tremblay}}]{kancharla08}%
  \BibitemOpen
  \bibfield  {author} {\bibinfo {author} {\bibfnamefont {S.~S.}\ \bibnamefont
  {Kancharla}}, \bibinfo {author} {\bibfnamefont {B.}~\bibnamefont {Kyung}},
  \bibinfo {author} {\bibfnamefont {D.}~\bibnamefont {S\'en\'echal}}, \bibinfo
  {author} {\bibfnamefont {M.}~\bibnamefont {Civelli}}, \bibinfo {author}
  {\bibfnamefont {M.}~\bibnamefont {Capone}}, \bibinfo {author} {\bibfnamefont
  {G.}~\bibnamefont {Kotliar}}, \ and\ \bibinfo {author} {\bibfnamefont
  {A.-M.~S.}\ \bibnamefont {Tremblay}},\ }\href {\doibase
  10.1103/PhysRevB.77.184516} {\bibfield  {journal} {\bibinfo  {journal} {Phys.
  Rev. B}\ }\textbf {\bibinfo {volume} {77}},\ \bibinfo {pages} {184516}
  (\bibinfo {year} {2008})}\BibitemShut {NoStop}%
\bibitem [{\citenamefont {Civelli}\ \emph {et~al.}(2008)\citenamefont
  {Civelli}, \citenamefont {Capone}, \citenamefont {Georges}, \citenamefont
  {Haule}, \citenamefont {Parcollet}, \citenamefont {Stanescu},\ and\
  \citenamefont {Kotliar}}]{civelli08}%
  \BibitemOpen
  \bibfield  {author} {\bibinfo {author} {\bibfnamefont {M.}~\bibnamefont
  {Civelli}}, \bibinfo {author} {\bibfnamefont {M.}~\bibnamefont {Capone}},
  \bibinfo {author} {\bibfnamefont {A.}~\bibnamefont {Georges}}, \bibinfo
  {author} {\bibfnamefont {K.}~\bibnamefont {Haule}}, \bibinfo {author}
  {\bibfnamefont {O.}~\bibnamefont {Parcollet}}, \bibinfo {author}
  {\bibfnamefont {T.}~\bibnamefont {Stanescu}}, \ and\ \bibinfo {author}
  {\bibfnamefont {G.}~\bibnamefont {Kotliar}},\ }\href@noop {} {\bibfield
  {journal} {\bibinfo  {journal} {Phys. Rev. Lett.}\ }\textbf {\bibinfo
  {volume} {100}},\ \bibinfo {pages} {046402} (\bibinfo {year}
  {2008})}\BibitemShut {NoStop}%
\bibitem [{\citenamefont {Kyung}\ \emph {et~al.}(2009)\citenamefont {Kyung},
  \citenamefont {S{\'e}n{\'e}chal},\ and\ \citenamefont {Tremblay}}]{kyung09}%
  \BibitemOpen
  \bibfield  {author} {\bibinfo {author} {\bibfnamefont {B.}~\bibnamefont
  {Kyung}}, \bibinfo {author} {\bibfnamefont {D.}~\bibnamefont
  {S{\'e}n{\'e}chal}}, \ and\ \bibinfo {author} {\bibfnamefont {A.-M.}\
  \bibnamefont {Tremblay}},\ }\href@noop {} {\bibfield  {journal} {\bibinfo
  {journal} {Phys. Rev. B}\ }\textbf {\bibinfo {volume} {80}},\ \bibinfo
  {pages} {205109} (\bibinfo {year} {2009})}\BibitemShut {NoStop}%
\bibitem [{\citenamefont {Civelli}(2009{\natexlab{a}})}]{civelli09PRL}%
  \BibitemOpen
  \bibfield  {author} {\bibinfo {author} {\bibfnamefont {M.}~\bibnamefont
  {Civelli}},\ }\href@noop {} {\bibfield  {journal} {\bibinfo  {journal} {Phys.
  Rev. Lett.}\ }\textbf {\bibinfo {volume} {103}},\ \bibinfo {pages} {136402}
  (\bibinfo {year} {2009}{\natexlab{a}})}\BibitemShut {NoStop}%
\bibitem [{\citenamefont {Civelli}(2009{\natexlab{b}})}]{civelli09PRB}%
  \BibitemOpen
  \bibfield  {author} {\bibinfo {author} {\bibfnamefont {M.}~\bibnamefont
  {Civelli}},\ }\href {\doibase 10.1103/PhysRevB.79.195113} {\bibfield
  {journal} {\bibinfo  {journal} {Phys. Rev. B}\ }\textbf {\bibinfo {volume}
  {79}},\ \bibinfo {pages} {195113} (\bibinfo {year}
  {2009}{\natexlab{b}})}\BibitemShut {NoStop}%
\bibitem [{\citenamefont {Sordi}\ \emph {et~al.}(2012)\citenamefont {Sordi},
  \citenamefont {S\'emon}, \citenamefont {Haule},\ and\ \citenamefont
  {Tremblay}}]{sordi12PRL}%
  \BibitemOpen
  \bibfield  {author} {\bibinfo {author} {\bibfnamefont {G.}~\bibnamefont
  {Sordi}}, \bibinfo {author} {\bibfnamefont {P.}~\bibnamefont {S\'emon}},
  \bibinfo {author} {\bibfnamefont {K.}~\bibnamefont {Haule}}, \ and\ \bibinfo
  {author} {\bibfnamefont {A.-M.~S.}\ \bibnamefont {Tremblay}},\ }\href
  {\doibase 10.1103/PhysRevLett.108.216401} {\bibfield  {journal} {\bibinfo
  {journal} {Phys. Rev. Lett.}\ }\textbf {\bibinfo {volume} {108}},\ \bibinfo
  {pages} {216401} (\bibinfo {year} {2012})}\BibitemShut {NoStop}%
\bibitem [{\citenamefont {Gull}\ \emph {et~al.}(2013)\citenamefont {Gull},
  \citenamefont {Parcollet},\ and\ \citenamefont {Millis}}]{gull13PRL}%
  \BibitemOpen
  \bibfield  {author} {\bibinfo {author} {\bibfnamefont {E.}~\bibnamefont
  {Gull}}, \bibinfo {author} {\bibfnamefont {O.}~\bibnamefont {Parcollet}}, \
  and\ \bibinfo {author} {\bibfnamefont {A.~J.}\ \bibnamefont {Millis}},\
  }\href {\doibase 10.1103/PhysRevLett.110.216405} {\bibfield  {journal}
  {\bibinfo  {journal} {Phys. Rev. Lett.}\ }\textbf {\bibinfo {volume} {110}},\
  \bibinfo {pages} {216405} (\bibinfo {year} {2013})}\BibitemShut {NoStop}%
\bibitem [{\citenamefont {Gull}\ and\ \citenamefont {Millis}(2015)}]{gull15}%
  \BibitemOpen
  \bibfield  {author} {\bibinfo {author} {\bibfnamefont {E.}~\bibnamefont
  {Gull}}\ and\ \bibinfo {author} {\bibfnamefont {A.~J.}\ \bibnamefont
  {Millis}},\ }\href {\doibase 10.1103/PhysRevB.91.085116} {\bibfield
  {journal} {\bibinfo  {journal} {Phys. Rev. B}\ }\textbf {\bibinfo {volume}
  {91}},\ \bibinfo {pages} {085116} (\bibinfo {year} {2015})}\BibitemShut
  {NoStop}%
\bibitem [{\citenamefont {Sakai}\ \emph {et~al.}(2016)\citenamefont {Sakai},
  \citenamefont {Civelli},\ and\ \citenamefont {Imada}}]{sakai16}%
  \BibitemOpen
  \bibfield  {author} {\bibinfo {author} {\bibfnamefont {S.}~\bibnamefont
  {Sakai}}, \bibinfo {author} {\bibfnamefont {M.}~\bibnamefont {Civelli}}, \
  and\ \bibinfo {author} {\bibfnamefont {M.}~\bibnamefont {Imada}},\ }\href
  {\doibase 10.1103/PhysRevLett.116.057003} {\bibfield  {journal} {\bibinfo
  {journal} {Phys. Rev. Lett.}\ }\textbf {\bibinfo {volume} {116}},\ \bibinfo
  {pages} {057003} (\bibinfo {year} {2016})}\BibitemShut {NoStop}%
\bibitem [{\citenamefont {{Harland}}\ \emph {et~al.}(2016)\citenamefont
  {{Harland}}, \citenamefont {{Katsnelson}},\ and\ \citenamefont
  {{Lichtenstein}}}]{harland16}%
  \BibitemOpen
  \bibfield  {author} {\bibinfo {author} {\bibfnamefont {M.}~\bibnamefont
  {{Harland}}}, \bibinfo {author} {\bibfnamefont {M.~I.}\ \bibnamefont
  {{Katsnelson}}}, \ and\ \bibinfo {author} {\bibfnamefont {A.~I.}\
  \bibnamefont {{Lichtenstein}}},\ }\href@noop {} {\bibfield  {journal}
  {\bibinfo  {journal} {ArXiv e-prints}\ } (\bibinfo {year} {2016})},\ \Eprint
  {http://arxiv.org/abs/1604.01808} {arXiv:1604.01808 [cond-mat.str-el]}
  \BibitemShut {NoStop}%
\bibitem [{\citenamefont {Huscroft}\ \emph {et~al.}(2001)\citenamefont
  {Huscroft}, \citenamefont {Jarrell}, \citenamefont {Maier}, \citenamefont
  {Moukouri},\ and\ \citenamefont {Tahvildarzadeh}}]{huscroft01}%
  \BibitemOpen
  \bibfield  {author} {\bibinfo {author} {\bibfnamefont {C.}~\bibnamefont
  {Huscroft}}, \bibinfo {author} {\bibfnamefont {M.}~\bibnamefont {Jarrell}},
  \bibinfo {author} {\bibfnamefont {T.}~\bibnamefont {Maier}}, \bibinfo
  {author} {\bibfnamefont {S.}~\bibnamefont {Moukouri}}, \ and\ \bibinfo
  {author} {\bibfnamefont {A.~N.}\ \bibnamefont {Tahvildarzadeh}},\ }\href
  {\doibase 10.1103/PhysRevLett.86.139} {\bibfield  {journal} {\bibinfo
  {journal} {Phys. Rev. Lett.}\ }\textbf {\bibinfo {volume} {86}},\ \bibinfo
  {pages} {139} (\bibinfo {year} {2001})}\BibitemShut {NoStop}%
\bibitem [{\citenamefont {Maier}\ \emph {et~al.}(2002)\citenamefont {Maier},
  \citenamefont {Pruschke},\ and\ \citenamefont {Jarrell}}]{maier02}%
  \BibitemOpen
  \bibfield  {author} {\bibinfo {author} {\bibfnamefont {T.~A.}\ \bibnamefont
  {Maier}}, \bibinfo {author} {\bibfnamefont {T.}~\bibnamefont {Pruschke}}, \
  and\ \bibinfo {author} {\bibfnamefont {M.}~\bibnamefont {Jarrell}},\
  }\href@noop {} {\bibfield  {journal} {\bibinfo  {journal} {Phys. Rev. B}\
  }\textbf {\bibinfo {volume} {66}},\ \bibinfo {pages} {075102} (\bibinfo
  {year} {2002})}\BibitemShut {NoStop}%
\bibitem [{\citenamefont {S\'en\'echal}\ and\ \citenamefont
  {Tremblay}(2004)}]{senechal04}%
  \BibitemOpen
  \bibfield  {author} {\bibinfo {author} {\bibfnamefont {D.}~\bibnamefont
  {S\'en\'echal}}\ and\ \bibinfo {author} {\bibfnamefont {A.-M.~S.}\
  \bibnamefont {Tremblay}},\ }\href {\doibase 10.1103/PhysRevLett.92.126401}
  {\bibfield  {journal} {\bibinfo  {journal} {Phys. Rev. Lett.}\ }\textbf
  {\bibinfo {volume} {92}},\ \bibinfo {pages} {126401} (\bibinfo {year}
  {2004})}\BibitemShut {NoStop}%
\bibitem [{\citenamefont {Civelli}\ \emph {et~al.}(2005)\citenamefont
  {Civelli}, \citenamefont {Capone}, \citenamefont {Kancharla}, \citenamefont
  {Parcollet},\ and\ \citenamefont {Kotliar}}]{civelli05}%
  \BibitemOpen
  \bibfield  {author} {\bibinfo {author} {\bibfnamefont {M.}~\bibnamefont
  {Civelli}}, \bibinfo {author} {\bibfnamefont {M.}~\bibnamefont {Capone}},
  \bibinfo {author} {\bibfnamefont {S.~S.}\ \bibnamefont {Kancharla}}, \bibinfo
  {author} {\bibfnamefont {O.}~\bibnamefont {Parcollet}}, \ and\ \bibinfo
  {author} {\bibfnamefont {G.}~\bibnamefont {Kotliar}},\ }\href {\doibase
  10.1103/PhysRevLett.95.106402} {\bibfield  {journal} {\bibinfo  {journal}
  {Phys. Rev. Lett.}\ }\textbf {\bibinfo {volume} {95}},\ \bibinfo {pages}
  {106402} (\bibinfo {year} {2005})}\BibitemShut {NoStop}%
\bibitem [{\citenamefont {Kyung}\ \emph {et~al.}(2006)\citenamefont {Kyung},
  \citenamefont {Kancharla}, \citenamefont {S\'en\'echal}, \citenamefont
  {Tremblay}, \citenamefont {Civelli},\ and\ \citenamefont
  {Kotliar}}]{kyung06PRB1}%
  \BibitemOpen
  \bibfield  {author} {\bibinfo {author} {\bibfnamefont {B.}~\bibnamefont
  {Kyung}}, \bibinfo {author} {\bibfnamefont {S.~S.}\ \bibnamefont
  {Kancharla}}, \bibinfo {author} {\bibfnamefont {D.}~\bibnamefont
  {S\'en\'echal}}, \bibinfo {author} {\bibfnamefont {A.-M.~S.}\ \bibnamefont
  {Tremblay}}, \bibinfo {author} {\bibfnamefont {M.}~\bibnamefont {Civelli}}, \
  and\ \bibinfo {author} {\bibfnamefont {G.}~\bibnamefont {Kotliar}},\ }\href
  {\doibase 10.1103/PhysRevB.73.165114} {\bibfield  {journal} {\bibinfo
  {journal} {Phys. Rev. B}\ }\textbf {\bibinfo {volume} {73}},\ \bibinfo
  {pages} {165114} (\bibinfo {year} {2006})}\BibitemShut {NoStop}%
\bibitem [{\citenamefont {Stanescu}\ and\ \citenamefont
  {Kotliar}(2006)}]{stanescu06}%
  \BibitemOpen
  \bibfield  {author} {\bibinfo {author} {\bibfnamefont {T.~D.}\ \bibnamefont
  {Stanescu}}\ and\ \bibinfo {author} {\bibfnamefont {G.}~\bibnamefont
  {Kotliar}},\ }\href {\doibase 10.1103/PhysRevB.74.125110} {\bibfield
  {journal} {\bibinfo  {journal} {Phys. Rev. B}\ }\textbf {\bibinfo {volume}
  {74}},\ \bibinfo {pages} {125110} (\bibinfo {year} {2006})}\BibitemShut
  {NoStop}%
\bibitem [{\citenamefont {Liebsch}\ and\ \citenamefont
  {Tong}(2009)}]{liebsch09}%
  \BibitemOpen
  \bibfield  {author} {\bibinfo {author} {\bibfnamefont {A.}~\bibnamefont
  {Liebsch}}\ and\ \bibinfo {author} {\bibfnamefont {N.-H.}\ \bibnamefont
  {Tong}},\ }\href {\doibase 10.1103/PhysRevB.80.165126} {\bibfield  {journal}
  {\bibinfo  {journal} {Phys. Rev. B}\ }\textbf {\bibinfo {volume} {80}},\
  \bibinfo {pages} {165126} (\bibinfo {year} {2009})}\BibitemShut {NoStop}%
\bibitem [{\citenamefont {Sakai}\ \emph
  {et~al.}(2009{\natexlab{a}})\citenamefont {Sakai}, \citenamefont {Motome},\
  and\ \citenamefont {Imada}}]{sakai09PRL}%
  \BibitemOpen
  \bibfield  {author} {\bibinfo {author} {\bibfnamefont {S.}~\bibnamefont
  {Sakai}}, \bibinfo {author} {\bibfnamefont {Y.}~\bibnamefont {Motome}}, \
  and\ \bibinfo {author} {\bibfnamefont {M.}~\bibnamefont {Imada}},\ }\href
  {\doibase 10.1103/PhysRevLett.102.056404} {\bibfield  {journal} {\bibinfo
  {journal} {Phys. Rev. Lett.}\ }\textbf {\bibinfo {volume} {102}},\ \bibinfo
  {pages} {056404} (\bibinfo {year} {2009}{\natexlab{a}})}\BibitemShut
  {NoStop}%
\bibitem [{\citenamefont {Sakai}\ \emph {et~al.}(2010)\citenamefont {Sakai},
  \citenamefont {Motome},\ and\ \citenamefont {Imada}}]{sakai10}%
  \BibitemOpen
  \bibfield  {author} {\bibinfo {author} {\bibfnamefont {S.}~\bibnamefont
  {Sakai}}, \bibinfo {author} {\bibfnamefont {Y.}~\bibnamefont {Motome}}, \
  and\ \bibinfo {author} {\bibfnamefont {M.}~\bibnamefont {Imada}},\ }\href
  {\doibase 10.1103/PhysRevB.82.134505} {\bibfield  {journal} {\bibinfo
  {journal} {Phys. Rev. B}\ }\textbf {\bibinfo {volume} {82}},\ \bibinfo
  {pages} {134505} (\bibinfo {year} {2010})}\BibitemShut {NoStop}%
\bibitem [{\citenamefont {Okamoto}\ \emph {et~al.}(2010)\citenamefont
  {Okamoto}, \citenamefont {Senechal}, \citenamefont {Civelli},\ and\
  \citenamefont {Tremblay}}]{okamoto10PRB82}%
  \BibitemOpen
  \bibfield  {author} {\bibinfo {author} {\bibfnamefont {S.}~\bibnamefont
  {Okamoto}}, \bibinfo {author} {\bibfnamefont {D.}~\bibnamefont {Senechal}},
  \bibinfo {author} {\bibfnamefont {M.}~\bibnamefont {Civelli}}, \ and\
  \bibinfo {author} {\bibfnamefont {A.-M.}\ \bibnamefont {Tremblay}},\
  }\href@noop {} {\bibfield  {journal} {\bibinfo  {journal} {Phys. Rev. B}\
  }\textbf {\bibinfo {volume} {82}},\ \bibinfo {pages} {180511} (\bibinfo
  {year} {2010})}\BibitemShut {NoStop}%
\bibitem [{\citenamefont {Eder}\ \emph {et~al.}(2011)\citenamefont {Eder},
  \citenamefont {Seki},\ and\ \citenamefont {Ohta}}]{eder11}%
  \BibitemOpen
  \bibfield  {author} {\bibinfo {author} {\bibfnamefont {R.}~\bibnamefont
  {Eder}}, \bibinfo {author} {\bibfnamefont {K.}~\bibnamefont {Seki}}, \ and\
  \bibinfo {author} {\bibfnamefont {Y.}~\bibnamefont {Ohta}},\ }\href {\doibase
  10.1103/PhysRevB.83.205137} {\bibfield  {journal} {\bibinfo  {journal} {Phys.
  Rev. B}\ }\textbf {\bibinfo {volume} {83}},\ \bibinfo {pages} {205137}
  (\bibinfo {year} {2011})}\BibitemShut {NoStop}%
\bibitem [{\citenamefont {Kohno}(2012)}]{kohno12}%
  \BibitemOpen
  \bibfield  {author} {\bibinfo {author} {\bibfnamefont {M.}~\bibnamefont
  {Kohno}},\ }\href {\doibase 10.1103/PhysRevLett.108.076401} {\bibfield
  {journal} {\bibinfo  {journal} {Phys. Rev. Lett.}\ }\textbf {\bibinfo
  {volume} {108}},\ \bibinfo {pages} {076401} (\bibinfo {year}
  {2012})}\BibitemShut {NoStop}%
\bibitem [{\citenamefont {Sakai}\ \emph {et~al.}(2013)\citenamefont {Sakai},
  \citenamefont {Blanc}, \citenamefont {Civelli}, \citenamefont {Gallais},
  \citenamefont {Cazayous}, \citenamefont {M\'easson}, \citenamefont {Wen},
  \citenamefont {Xu}, \citenamefont {Gu}, \citenamefont {Sangiovanni},
  \citenamefont {Motome}, \citenamefont {Held}, \citenamefont {Sacuto},
  \citenamefont {Georges},\ and\ \citenamefont {Imada}}]{sakai13}%
  \BibitemOpen
  \bibfield  {author} {\bibinfo {author} {\bibfnamefont {S.}~\bibnamefont
  {Sakai}}, \bibinfo {author} {\bibfnamefont {S.}~\bibnamefont {Blanc}},
  \bibinfo {author} {\bibfnamefont {M.}~\bibnamefont {Civelli}}, \bibinfo
  {author} {\bibfnamefont {Y.}~\bibnamefont {Gallais}}, \bibinfo {author}
  {\bibfnamefont {M.}~\bibnamefont {Cazayous}}, \bibinfo {author}
  {\bibfnamefont {M.-A.}\ \bibnamefont {M\'easson}}, \bibinfo {author}
  {\bibfnamefont {J.~S.}\ \bibnamefont {Wen}}, \bibinfo {author} {\bibfnamefont
  {Z.~J.}\ \bibnamefont {Xu}}, \bibinfo {author} {\bibfnamefont {G.~D.}\
  \bibnamefont {Gu}}, \bibinfo {author} {\bibfnamefont {G.}~\bibnamefont
  {Sangiovanni}}, \bibinfo {author} {\bibfnamefont {Y.}~\bibnamefont {Motome}},
  \bibinfo {author} {\bibfnamefont {K.}~\bibnamefont {Held}}, \bibinfo {author}
  {\bibfnamefont {A.}~\bibnamefont {Sacuto}}, \bibinfo {author} {\bibfnamefont
  {A.}~\bibnamefont {Georges}}, \ and\ \bibinfo {author} {\bibfnamefont
  {M.}~\bibnamefont {Imada}},\ }\href {\doibase 10.1103/PhysRevLett.111.107001}
  {\bibfield  {journal} {\bibinfo  {journal} {Phys. Rev. Lett.}\ }\textbf
  {\bibinfo {volume} {111}},\ \bibinfo {pages} {107001} (\bibinfo {year}
  {2013})}\BibitemShut {NoStop}%
\bibitem [{\citenamefont {Kohno}(2014)}]{kohno14}%
  \BibitemOpen
  \bibfield  {author} {\bibinfo {author} {\bibfnamefont {M.}~\bibnamefont
  {Kohno}},\ }\href {\doibase 10.1103/PhysRevB.90.035111} {\bibfield  {journal}
  {\bibinfo  {journal} {Phys. Rev. B}\ }\textbf {\bibinfo {volume} {90}},\
  \bibinfo {pages} {035111} (\bibinfo {year} {2014})}\BibitemShut {NoStop}%
\bibitem [{\citenamefont {Merino}\ \emph {et~al.}(2016)\citenamefont {Merino},
  \citenamefont {Gunnarsson},\ and\ \citenamefont {Kotliar}}]{merino16}%
  \BibitemOpen
  \bibfield  {author} {\bibinfo {author} {\bibfnamefont {J.}~\bibnamefont
  {Merino}}, \bibinfo {author} {\bibfnamefont {O.}~\bibnamefont {Gunnarsson}},
  \ and\ \bibinfo {author} {\bibfnamefont {G.}~\bibnamefont {Kotliar}},\ }\href
  {http://stacks.iop.org/0953-8984/28/i=4/a=045501} {\bibfield  {journal}
  {\bibinfo  {journal} {J.Phys.: Condens. Matter}\ }\textbf {\bibinfo {volume}
  {28}},\ \bibinfo {pages} {045501} (\bibinfo {year} {2016})}\BibitemShut
  {NoStop}%
\bibitem [{\citenamefont {Yang}\ and\ \citenamefont {Feiguin}(2016)}]{yang16}%
  \BibitemOpen
  \bibfield  {author} {\bibinfo {author} {\bibfnamefont {C.}~\bibnamefont
  {Yang}}\ and\ \bibinfo {author} {\bibfnamefont {A.~E.}\ \bibnamefont
  {Feiguin}},\ }\href {\doibase 10.1103/PhysRevB.93.081107} {\bibfield
  {journal} {\bibinfo  {journal} {Phys. Rev. B}\ }\textbf {\bibinfo {volume}
  {93}},\ \bibinfo {pages} {081107} (\bibinfo {year} {2016})}\BibitemShut
  {NoStop}%
\bibitem [{\citenamefont {Maier}\ \emph {et~al.}(2008)\citenamefont {Maier},
  \citenamefont {Poilblanc},\ and\ \citenamefont {Scalapino}}]{maier08}%
  \BibitemOpen
  \bibfield  {author} {\bibinfo {author} {\bibfnamefont {T.~A.}\ \bibnamefont
  {Maier}}, \bibinfo {author} {\bibfnamefont {D.}~\bibnamefont {Poilblanc}}, \
  and\ \bibinfo {author} {\bibfnamefont {D.~J.}\ \bibnamefont {Scalapino}},\
  }\href {\doibase 10.1103/PhysRevLett.100.237001} {\bibfield  {journal}
  {\bibinfo  {journal} {Phys. Rev. Lett.}\ }\textbf {\bibinfo {volume} {100}},\
  \bibinfo {pages} {237001} (\bibinfo {year} {2008})}\BibitemShut {NoStop}%
\bibitem [{\citenamefont {Luttinger}(1961)}]{luttinger61}%
  \BibitemOpen
  \bibfield  {author} {\bibinfo {author} {\bibfnamefont {J.~M.}\ \bibnamefont
  {Luttinger}},\ }\href {\doibase 10.1103/PhysRev.121.942} {\bibfield
  {journal} {\bibinfo  {journal} {Phys. Rev.}\ }\textbf {\bibinfo {volume}
  {121}},\ \bibinfo {pages} {942} (\bibinfo {year} {1961})}\BibitemShut
  {NoStop}%
\bibitem [{\citenamefont {Hybertsen}\ \emph {et~al.}(1990)\citenamefont
  {Hybertsen}, \citenamefont {Stechel}, \citenamefont {Schluter},\ and\
  \citenamefont {Jennison}}]{hybertsen90}%
  \BibitemOpen
  \bibfield  {author} {\bibinfo {author} {\bibfnamefont {M.~S.}\ \bibnamefont
  {Hybertsen}}, \bibinfo {author} {\bibfnamefont {E.}~\bibnamefont {Stechel}},
  \bibinfo {author} {\bibfnamefont {M.}~\bibnamefont {Schluter}}, \ and\
  \bibinfo {author} {\bibfnamefont {D.}~\bibnamefont {Jennison}},\ }\href@noop
  {} {\bibfield  {journal} {\bibinfo  {journal} {Phys. Rev. B}\ }\textbf
  {\bibinfo {volume} {41}},\ \bibinfo {pages} {11068} (\bibinfo {year}
  {1990})}\BibitemShut {NoStop}%
\bibitem [{\citenamefont {Capone}\ \emph {et~al.}(2007)\citenamefont {Capone},
  \citenamefont {de' Medici},\ and\ \citenamefont {Georges}}]{capone07}%
  \BibitemOpen
  \bibfield  {author} {\bibinfo {author} {\bibfnamefont {M.}~\bibnamefont
  {Capone}}, \bibinfo {author} {\bibfnamefont {L.}~\bibnamefont {de' Medici}},
  \ and\ \bibinfo {author} {\bibfnamefont {A.}~\bibnamefont {Georges}},\ }\href
  {\doibase 10.1103/PhysRevB.76.245116} {\bibfield  {journal} {\bibinfo
  {journal} {Phys. Rev. B}\ }\textbf {\bibinfo {volume} {76}},\ \bibinfo
  {pages} {245116} (\bibinfo {year} {2007})}\BibitemShut {NoStop}%
\bibitem [{\citenamefont {Liebsch}\ \emph {et~al.}(2008)\citenamefont
  {Liebsch}, \citenamefont {Ishida},\ and\ \citenamefont {Merino}}]{liebsch08}%
  \BibitemOpen
  \bibfield  {author} {\bibinfo {author} {\bibfnamefont {A.}~\bibnamefont
  {Liebsch}}, \bibinfo {author} {\bibfnamefont {H.}~\bibnamefont {Ishida}}, \
  and\ \bibinfo {author} {\bibfnamefont {J.}~\bibnamefont {Merino}},\
  }\href@noop {} {\bibfield  {journal} {\bibinfo  {journal} {Phys. Rev. B}\
  }\textbf {\bibinfo {volume} {78}},\ \bibinfo {pages} {165123} (\bibinfo
  {year} {2008})}\BibitemShut {NoStop}%
\bibitem [{\citenamefont {Liebsch}\ and\ \citenamefont
  {Ishida}(2012)}]{liebsch12}%
  \BibitemOpen
  \bibfield  {author} {\bibinfo {author} {\bibfnamefont {A.}~\bibnamefont
  {Liebsch}}\ and\ \bibinfo {author} {\bibfnamefont {H.}~\bibnamefont
  {Ishida}},\ }\href {http://stacks.iop.org/0953-8984/24/i=5/a=053201}
  {\bibfield  {journal} {\bibinfo  {journal} {J. Phys.: Condens. Matter}\
  }\textbf {\bibinfo {volume} {24}},\ \bibinfo {pages} {053201} (\bibinfo
  {year} {2012})}\BibitemShut {NoStop}%
\bibitem [{Note1()}]{Note1}%
  \BibitemOpen
  \bibinfo {note} {This part for the normal state is essentially equivalent to
  what has been done in a recent paper Ref.~\protect \rev@citealpnum
  {seki16}.}\BibitemShut {Stop}%
\bibitem [{\citenamefont {Seki}\ and\ \citenamefont {Yunoki}(2016)}]{seki16}%
  \BibitemOpen
  \bibfield  {author} {\bibinfo {author} {\bibfnamefont {K.}~\bibnamefont
  {Seki}}\ and\ \bibinfo {author} {\bibfnamefont {S.}~\bibnamefont {Yunoki}},\
  }\href@noop {} {\bibfield  {journal} {\bibinfo  {journal} {ArXiv e-prints}\ }
  (\bibinfo {year} {2016})},\ \Eprint {http://arxiv.org/abs/1603.07794}
  {arXiv:1603.07794 [cond-mat.str-el]} \BibitemShut {NoStop}%
\bibitem [{\citenamefont {Seki}\ \emph {et~al.}(2011)\citenamefont {Seki},
  \citenamefont {Eder},\ and\ \citenamefont {Ohta}}]{seki11}%
  \BibitemOpen
  \bibfield  {author} {\bibinfo {author} {\bibfnamefont {K.}~\bibnamefont
  {Seki}}, \bibinfo {author} {\bibfnamefont {R.}~\bibnamefont {Eder}}, \ and\
  \bibinfo {author} {\bibfnamefont {Y.}~\bibnamefont {Ohta}},\ }\href {\doibase
  10.1103/PhysRevB.84.245106} {\bibfield  {journal} {\bibinfo  {journal} {Phys.
  Rev. B}\ }\textbf {\bibinfo {volume} {84}},\ \bibinfo {pages} {245106}
  (\bibinfo {year} {2011})}\BibitemShut {NoStop}%
\bibitem [{\citenamefont {Sakai}\ \emph
  {et~al.}(2009{\natexlab{b}})\citenamefont {Sakai}, \citenamefont {Motome},\
  and\ \citenamefont {Imada}}]{sakai09PhysB}%
  \BibitemOpen
  \bibfield  {author} {\bibinfo {author} {\bibfnamefont {S.}~\bibnamefont
  {Sakai}}, \bibinfo {author} {\bibfnamefont {Y.}~\bibnamefont {Motome}}, \
  and\ \bibinfo {author} {\bibfnamefont {M.}~\bibnamefont {Imada}},\
  }\href@noop {} {\bibfield  {journal} {\bibinfo  {journal} {Physica B:
  Condensed Matter}\ }\textbf {\bibinfo {volume} {404}},\ \bibinfo {pages}
  {3183} (\bibinfo {year} {2009}{\natexlab{b}})}\BibitemShut {NoStop}%
\bibitem [{\citenamefont {Zhang}\ and\ \citenamefont {Imada}(2007)}]{zhang07}%
  \BibitemOpen
  \bibfield  {author} {\bibinfo {author} {\bibfnamefont {Y.~Z.}\ \bibnamefont
  {Zhang}}\ and\ \bibinfo {author} {\bibfnamefont {M.}~\bibnamefont {Imada}},\
  }\href {\doibase 10.1103/PhysRevB.76.045108} {\bibfield  {journal} {\bibinfo
  {journal} {Phys. Rev. B}\ }\textbf {\bibinfo {volume} {76}},\ \bibinfo
  {pages} {045108} (\bibinfo {year} {2007})}\BibitemShut {NoStop}%
\bibitem [{\citenamefont {Park}\ \emph {et~al.}(2008)\citenamefont {Park},
  \citenamefont {Haule},\ and\ \citenamefont {Kotliar}}]{park08}%
  \BibitemOpen
  \bibfield  {author} {\bibinfo {author} {\bibfnamefont {H.}~\bibnamefont
  {Park}}, \bibinfo {author} {\bibfnamefont {K.}~\bibnamefont {Haule}}, \ and\
  \bibinfo {author} {\bibfnamefont {G.}~\bibnamefont {Kotliar}},\ }\href
  {\doibase 10.1103/PhysRevLett.101.186403} {\bibfield  {journal} {\bibinfo
  {journal} {Phys. Rev. Lett.}\ }\textbf {\bibinfo {volume} {101}},\ \bibinfo
  {pages} {186403} (\bibinfo {year} {2008})}\BibitemShut {NoStop}%
\bibitem [{\citenamefont {Loret}\ \emph {et~al.}(2016)\citenamefont {Loret},
  \citenamefont {Sakai}, \citenamefont {Gallais}, \citenamefont {Cazayous},
  \citenamefont {M\'easson}, \citenamefont {Forget}, \citenamefont {Colson},
  \citenamefont {Civelli},\ and\ \citenamefont {Sacuto}}]{loret16}%
  \BibitemOpen
  \bibfield  {author} {\bibinfo {author} {\bibfnamefont {B.}~\bibnamefont
  {Loret}}, \bibinfo {author} {\bibfnamefont {S.}~\bibnamefont {Sakai}},
  \bibinfo {author} {\bibfnamefont {Y.}~\bibnamefont {Gallais}}, \bibinfo
  {author} {\bibfnamefont {M.}~\bibnamefont {Cazayous}}, \bibinfo {author}
  {\bibfnamefont {M.-A.}\ \bibnamefont {M\'easson}}, \bibinfo {author}
  {\bibfnamefont {A.}~\bibnamefont {Forget}}, \bibinfo {author} {\bibfnamefont
  {D.}~\bibnamefont {Colson}}, \bibinfo {author} {\bibfnamefont
  {M.}~\bibnamefont {Civelli}}, \ and\ \bibinfo {author} {\bibfnamefont
  {A.}~\bibnamefont {Sacuto}},\ }\href {\doibase
  10.1103/PhysRevLett.116.197001} {\bibfield  {journal} {\bibinfo  {journal}
  {Phys. Rev. Lett.}\ }\textbf {\bibinfo {volume} {116}},\ \bibinfo {pages}
  {197001} (\bibinfo {year} {2016})}\BibitemShut {NoStop}%
\bibitem [{Note2()}]{Note2}%
  \BibitemOpen
  \bibinfo {note} {Note that the sign of $D_{f_1}$ depends on which of ($\pi
  ,0$) and (0,$\pi $) we choose to plot for a $d$-wave symmetry-broken state
  and does not matter.}\BibitemShut {Stop}%
\bibitem [{Note3()}]{Note3}%
  \BibitemOpen
  \bibinfo {note} {This is true even when we consider the superconducting gap
  function $\Delta $ rather than $\Sigma ^{\protect \rm ano}$ because $\Delta $
  in the hidden-fermion model has a pole at $\omega =\pm \protect \sqrt
  {\epsilon _{f_1}^2+D_{f_1}^2+V_{1}^2}$ \cite {sakai16}}\BibitemShut {NoStop}%
\bibitem [{\citenamefont {Pushp}\ \emph {et~al.}(2009)\citenamefont {Pushp},
  \citenamefont {Parker}, \citenamefont {Pasupathy}, \citenamefont {Gomes},
  \citenamefont {Ono}, \citenamefont {Wen}, \citenamefont {Xu}, \citenamefont
  {Gu},\ and\ \citenamefont {Yazdani}}]{pushp09}%
  \BibitemOpen
  \bibfield  {author} {\bibinfo {author} {\bibfnamefont {A.}~\bibnamefont
  {Pushp}}, \bibinfo {author} {\bibfnamefont {C.~V.}\ \bibnamefont {Parker}},
  \bibinfo {author} {\bibfnamefont {A.~N.}\ \bibnamefont {Pasupathy}}, \bibinfo
  {author} {\bibfnamefont {K.~K.}\ \bibnamefont {Gomes}}, \bibinfo {author}
  {\bibfnamefont {S.}~\bibnamefont {Ono}}, \bibinfo {author} {\bibfnamefont
  {J.}~\bibnamefont {Wen}}, \bibinfo {author} {\bibfnamefont {Z.}~\bibnamefont
  {Xu}}, \bibinfo {author} {\bibfnamefont {G.}~\bibnamefont {Gu}}, \ and\
  \bibinfo {author} {\bibfnamefont {A.}~\bibnamefont {Yazdani}},\ }\href
  {\doibase 10.1126/science.1174338} {\bibfield  {journal} {\bibinfo  {journal}
  {Science}\ } (\bibinfo {year} {2009}),\ 10.1126/science.1174338}\BibitemShut
  {NoStop}%
\bibitem [{\citenamefont {Sakai}\ \emph {et~al.}(2015)\citenamefont {Sakai},
  \citenamefont {Civelli}, \citenamefont {Nomura},\ and\ \citenamefont
  {Imada}}]{sakai15}%
  \BibitemOpen
  \bibfield  {author} {\bibinfo {author} {\bibfnamefont {S.}~\bibnamefont
  {Sakai}}, \bibinfo {author} {\bibfnamefont {M.}~\bibnamefont {Civelli}},
  \bibinfo {author} {\bibfnamefont {Y.}~\bibnamefont {Nomura}}, \ and\ \bibinfo
  {author} {\bibfnamefont {M.}~\bibnamefont {Imada}},\ }\href@noop {}
  {\bibfield  {journal} {\bibinfo  {journal} {Phys. Rev. B}\ }\textbf {\bibinfo
  {volume} {92}},\ \bibinfo {pages} {180503} (\bibinfo {year}
  {2015})}\BibitemShut {NoStop}%
\bibitem [{\citenamefont {Yamaji}\ and\ \citenamefont
  {Imada}(2011)}]{yamaji11PRL}%
  \BibitemOpen
  \bibfield  {author} {\bibinfo {author} {\bibfnamefont {Y.}~\bibnamefont
  {Yamaji}}\ and\ \bibinfo {author} {\bibfnamefont {M.}~\bibnamefont {Imada}},\
  }\href@noop {} {\bibfield  {journal} {\bibinfo  {journal} {Phys. Rev. Lett.}\
  }\textbf {\bibinfo {volume} {106}},\ \bibinfo {pages} {016404} (\bibinfo
  {year} {2011})}\BibitemShut {NoStop}%
\bibitem [{\citenamefont {Misawa}\ and\ \citenamefont
  {Imada}(2014)}]{misawa14}%
  \BibitemOpen
  \bibfield  {author} {\bibinfo {author} {\bibfnamefont {T.}~\bibnamefont
  {Misawa}}\ and\ \bibinfo {author} {\bibfnamefont {M.}~\bibnamefont {Imada}},\
  }\href {\doibase 10.1103/PhysRevB.90.115137} {\bibfield  {journal} {\bibinfo
  {journal} {Phys. Rev. B}\ }\textbf {\bibinfo {volume} {90}},\ \bibinfo
  {pages} {115137} (\bibinfo {year} {2014})}\BibitemShut {NoStop}%
\bibitem [{Note4()}]{Note4}%
  \BibitemOpen
  \bibinfo {note} {A caution is needed to discuss two or more electron
  processes within the hidden-fermion Hamiltonian because the vertex
  corrections in the original many-body system are not straightforwardly taken
  into account in the hidden-fermion Hamiltonian.}\BibitemShut {Stop}%
\bibitem [{\citenamefont {Yang}\ \emph {et~al.}(2006)\citenamefont {Yang},
  \citenamefont {Rice},\ and\ \citenamefont {Zhang}}]{yang06}%
  \BibitemOpen
  \bibfield  {author} {\bibinfo {author} {\bibfnamefont {K.-Y.}\ \bibnamefont
  {Yang}}, \bibinfo {author} {\bibfnamefont {T.}~\bibnamefont {Rice}}, \ and\
  \bibinfo {author} {\bibfnamefont {F.-C.}\ \bibnamefont {Zhang}},\ }\href@noop
  {} {\bibfield  {journal} {\bibinfo  {journal} {Phys. Rev. B}\ }\textbf
  {\bibinfo {volume} {73}},\ \bibinfo {pages} {174501} (\bibinfo {year}
  {2006})}\BibitemShut {NoStop}%
\bibitem [{\citenamefont {Rice}\ \emph {et~al.}(2012)\citenamefont {Rice},
  \citenamefont {Yang},\ and\ \citenamefont {Zhang}}]{rice12}%
  \BibitemOpen
  \bibfield  {author} {\bibinfo {author} {\bibfnamefont {T.~M.}\ \bibnamefont
  {Rice}}, \bibinfo {author} {\bibfnamefont {K.-Y.}\ \bibnamefont {Yang}}, \
  and\ \bibinfo {author} {\bibfnamefont {F.~C.}\ \bibnamefont {Zhang}},\ }\href
  {http://stacks.iop.org/0034-4885/75/i=1/a=016502} {\bibfield  {journal}
  {\bibinfo  {journal} {Rep. Prog. Phys.}\ }\textbf {\bibinfo {volume} {75}},\
  \bibinfo {pages} {016502} (\bibinfo {year} {2012})}\BibitemShut {NoStop}%
\bibitem [{Note5()}]{Note5}%
  \BibitemOpen
  \bibinfo {note} {The $\omega $-independent term $s(\protect \mathbf k)$ in
  Eq.~(\ref {sigc_sc}) should be included in the dispersion term of $c$, as is
  expressed in the first term of Eq.~(\ref {tcf_ns}).}\BibitemShut {Stop}%
\bibitem [{\citenamefont {Harris}\ and\ \citenamefont
  {Lange}(1967)}]{harris67}%
  \BibitemOpen
  \bibfield  {author} {\bibinfo {author} {\bibfnamefont {A.~B.}\ \bibnamefont
  {Harris}}\ and\ \bibinfo {author} {\bibfnamefont {R.~V.}\ \bibnamefont
  {Lange}},\ }\href {\doibase 10.1103/PhysRev.157.295} {\bibfield  {journal}
  {\bibinfo  {journal} {Phys. Rev.}\ }\textbf {\bibinfo {volume} {157}},\
  \bibinfo {pages} {295} (\bibinfo {year} {1967})}\BibitemShut {NoStop}%
\end{thebibliography}%

\end{document}